\documentclass[particles,article,accept,moreauthors,pdftex]{Definitions/mdpi}
\usepackage{amssymb,graphicx,bm,mathrsfs,color,amsmath,slashed,comment}
\newcommand{\bea}{\begin{eqnarray}}
\newcommand{\eea}{\end{eqnarray}}
\newcommand{\be}{\begin{equation}}
\newcommand{\ee}{\end{equation}}

\newcommand{\D}{\Delta}
\input{acronym.input}
\firstpage{1} 
\doinum{}
\pubvolume{}
\issuenum{}
\articlenumber{}
\pubyear{2020}
\copyrightyear{}
\history{}
\updates{} % If there is an update available, un-comment this line

\Title{ Equation of State of Strongly Magnetized Matter With Hyperons
  and $\Delta$-Resonances}
\Author{
  {Vivek Baruah Thapa} $^{1,}$*\orcidA{}, {{Monika Sinha}}%Please confirm the two corresponding authors on word are right,  Prof. Armen Sedrakian is the only corresponding author in SuSy system.
   $^{1,}$*\orcidB{}, Jia Jie Li $^{2,3}$\orcidC{} and Armen Sedrakian $^{4,5}$\orcidD{}
            }
\AuthorNames{Vivek Baruah Thapa, Monika Sinha, Jia Jie Li and Armen Sedrakian}
\address{\noindent
$^{1}$ \quad {Indian Institute of Technology Jodhpur, Jodhpur 342037, India} \\
$^{2}$ \quad {Institute of Modern Physics, Chinese Academy of Sciences,
Lanzhou 730000, China}; jiajieli@itp.uni-frankfurt.de\\
$^{3}$ \quad {Institute for Theoretical Physics, Goethe University, Max-von-Laue-Stra\ss e,~1, 60438~Frankfurt~am~Main,~Germany} \\
$^{4}$ \quad  Frankfurt Institute for Advanced Studies,
Ruth-Moufang-Stra\ss e, 1, 60438 Frankfurt am Main, Germany; sedrakian@fias.uni-frankfurt.de\\
$^{5}$ \quad  Institute of Theoretical Physics, University of
Wroc\l{}aw, pl. M. Borna 9, 50-204 Wroc\l{}aw, Poland \\
}

\corres{Correspondence: thapa.1@iitj.ac.in (V.B.T.); ms@iitj.ac.in (M.S.)}
\abstract{We construct a new equation of state for the baryonic matter
  under an intense magnetic field within the framework of covariant
  density functional theory. The composition of matter includes
  hyperons as well as $\D$-resonances.  The extension of the nucleonic
  functional to the hypernuclear sector is constrained by the
  experimental data on $\Lambda$ and $\Xi$-hypernuclei. We find that
  the equation of state stiffens with the inclusion of the magnetic
  field, which increases the maximum mass of neutron star compared to
  the non-magnetic case. In addition, the strangeness fraction in the
  matter is enhanced. Several observables, like the Dirac effective
  mass, particle abundances, etc. show typical oscillatory behavior as
  a function of the magnetic field and/or density which is traced back
  to the occupation pattern of Landau levels. }
\begin{document}

\section{Introduction}\label{intro}

Compact stars are the end products of stellar evolution that are
produced in supernova explosions. They are among the most fascinating
objects in the universe that motivate theoretical studies of exotic
states of matter which are difficult or impossible to produce in
modern terrestrial laboratories. Among~the remarkable features of
compact stars are the wide range of densities covered by their
interiors (from sub-saturation up to possibly 10 times the nuclear
saturation density) and the huge magnetic field range $10^9$ to
$10^{18}$ G.  The~compact stars are arranged in various classes
according to some of their characteristic features. These include
millisecond pulsars, neutron stars in X-ray binaries, radio pulsars,
anomalous X-ray pulsars, soft gamma repeaters, etc.  Among~these, the~anomalous X-ray pulsars and soft gamma repeaters are believed to be
compact stars with the surface magnetic field in the range of
$10^{14}$--$10^{15}$ G~\cite{Harding2006,Turolla2015RPPh} and are
commonly referred as {\it {magnetars}%Is the italics necessary?
}.  Furthermore, there~has been
recently growing evidence that (at least) the repeating fast radio
bursts (FRBs) are related to
magnetars~\cite{Margalit2020ApJ,Beniamini2020MNRAS,Beloborodov2020,Levin2020ApJ,Zanazzi2020ApJ}.
Since the gravitational equilibrium of compact stars admits stars with
magnetic fields in the range $B \leq 10^{18}$--$10^{19}$~G, large
magnetic fields beyond those currently inferred have been studied
theoretically. Earlier
works~\cite{1997PhRvL..78.2898C,1997PhRvL..79.2176B,2000ApJ...537..351B,2007MPLA...22..623C,2008JPhG...35l5201R}
have studied the effects of the magnetic field on the gross parameters
of compact stars, such as the mass, radius, moment of inertia under
intense magnetic fields. The~induced deformations of the neutron stars
(NSs) due to the strong magnetic fields can be important sources of
gravitational waves and precession in neutron stars, see
Refs.~\cite{2014PhRvC..89d5805M,1993A&A...278..421B,1995A&A...301..757B,2001ApJ...554..322C}. The~structure of magnetized compact stars, in~particular, their
deformation, in~general relativity, has been studied initially in
Refs.~\cite{1993A&A...278..421B,1995A&A...301..757B,2001ApJ...554..322C}
assuming various forms of the poloidal and toroidal field
configurations.  More recent
studies~\cite{2009MNRAS.397..913C,2010MNRAS.406.2540C,2013MNRAS.435L..43C}
considered also a combination of toroidal and poloidal fields which
appear to be more stable than purely poloidal or toroidal
configurations. Moreover, a~``universal'' field profile represented by
an 8th-order polynomial as a function of star's internal radius has
been proposed recently to describe the magnetic field profile inside
the star~\cite{2019PhRvC..99e5811C} based on the solution of
Einstein--Maxwell equations in general relativity.  While large
magnetic fields are required to affect the equation of state
{(i.e., the~dependence of pressure on the energy density, hereafter
  abbreviated as EoS)}
and the structure of the star, the~role of the
magnetic field is still important at lower values. In~particular,
MeV-scale magnetic field can significantly alter the quasiparticle
spectrum of baryons, leading to the suppression of the superfluidity
of protons via Landau diamagnetism~\cite{Sinha2014,Sinha2015} and
superfluidity of neutrons via Pauli paramagnetism~\cite{Stein2016},
see for a review~\cite{Sedrakian2017}. These modifications alter the
neutrino emissivity of compact stars with MeV-scale magnetic
fields~\cite{Sinha2015} through the modifications of the neutrino
production reaction rates. The~anisotropy introduced by the magnetic
field also affects evolutionary processes in compact stars such as
their magneto-thermal evolution~\cite{2009A&A...496..207P} and
rotational dynamics~\cite{2019LRCA....5....3P,Sedrakian2016}.

Recent observations of compact stars in a wide range of
electromagnetic spectra and in gravitational waves motivate detailed
microscopic studies of the interior matter, in~particular, its~EoS and
composition.  A~fundamental observational property of compact stars is
the maximum mass, which is still a matter of debate.  The~mass of PSR
1913+16 (the Hulse--Taylor pulsar) $1.4398$~M$_\odot$ is one of the
precisely determined pulsar masses~\cite{1975ApJ...195L..51H}. The~largest masses were measured for millisecond pulsars in binaries with
white dwarfs, namely J1614$-$2230 ($1.97\pm0.04$ M$_\odot$)
\cite{2010ApJ...724L.199O}, PSR~J0348+0432 \mbox{($2.01\pm 0.04$ M$_\odot$)}~\cite{2013Sci...340..448A} and MSP %please define PSR and MSP if appropriate, and please confirm the number, such as J0348+0432, is right. 
J0740+6620 ($2.14^{+0.20}_{-0.18}$
M$_\odot$ with 95$\%$ credibility)~\cite{2020NatAs...4...72C}. The~last measurement, which is based on Shapiro delay, is so far the
largest measured maximum mass with relatively small error bars and,
thus, sets a reliable lower bound on the maximum mass of a compact
object. Another recent observation of gravitational waves by the
LIGO-Virgo %please define if appropirate. 
Collaboration  (the ``GW190814''event) from a binary
system of a black hole and light compact object companion sets the mass
of the latter at $2.59^{+0.08}_{-0.09}$ M$_\odot$
\cite{2020ApJ...896L..44A}, but~the origin of this object (i.e., a~light black hole or a heavy neutron star) is not settled.  In~addition, high precision mass and radius measurements for the pulsar
PSR J0030+0451 are offered by the Neutron star Interior
Composition ExploreR (NICER) space mission with mass--radius values
$1.44^{+0.15}_{-0.14}$ M$_\odot$, $13.02^{+1.24}_{-1.06}$ km
~\cite{2019ApJ...887L..24M} and $1.34^{+0.15}_{-0.16}$ M$_\odot$,
$12.71^{+1.14}_{-1.19}$ km~\cite{2019ApJ...887L..21R}.

The composition of matter at about several times the nuclear saturation
density is not known. One possibility is that matter is nucleonic (with
a small admixture of leptons to guarantee the charge neutrality) up
to the center of a star. However, in~massive compact stars, the~densities can reach values exceeding the saturation density by an
order of magnitude. Therefore, the~appearance of new degrees of
freedom is a possibility. One option is the nucleation of hyperons,
which softens the EoS and results (for some models)
in a maximum mass of a compact star below the value observational
minimum~$2M_\odot$. There are several modern covariant density
functional (CDF)-based models which avoid this problem and provide
sufficient repulsion to produce stars with large enough masses $M >
2M_\odot$. A~particular class of these models, which we will use in
this work, is based on density functionals with density-dependent (DD)
couplings~\cite{1995PhLB..345..355L,1995PhRvC..52.3043F,1999NuPhA.656..331T,
2001PhRvC..64b5804H,2008PhR...464..113L}. The~interactions in these models are mediated via
the exchange of $\sigma$, $\omega$, $\rho$-mesons, and~in the hypernuclear
sector also via two additional (hidden strangeness) $\sigma^*$ and
$\phi$-mesons.

An interesting possibility is an appearance  of $\Delta$-resonances in
compact stars, which has regained attention in recent
years~\cite{2014PhRvD..89d3014D,2014PhRvC..90f5809D,2015PhRvC..92a5802C},
after they have been neglected for a long time due to presumed high
onset density of the order of $10$ times the nuclear saturation
density~\cite{1985ApJ...293..470G}.  The~strong interactions, in~this
case, are mediated by the exchange of $\sigma$, $\omega$ and~$\rho$-mesons only. The~inclusion of $\Delta$-resonances in the EoS shifts the
onset of hyperons to higher densities. Consequently, the~high-density
part of the EoS is stiffer and the maximum mass is higher when
$\Delta$-resonances are included. They also significantly reduce the
radius of the star~\cite{2018PhLB..783..234L,Li2019ApJ,2020arXiv200709683S} due
to the softening of the EoS at intermediate~densities.

In this work, we consider $\Delta$-resonance admixed hypernuclear
matter in strong magnetic fields within the DDME2 parameterization,
which has been used already for the same problem in the case of zero
magnetic fields in Refs.~\cite{2018PhLB..783..234L,Li2019ApJ,2019PhRvC.100a5809L,2020arXiv200709683S}.

The paper is organized as follows. In~Section~\ref{sec:formalism}, we
briefly discuss the CDF formalism in the presence of strong magnetic
fields. Our results are shown in Section~\ref{sec:results} and our
conclusions are collected in Section~\ref{sec:conclusions}.

\section{Formalism}
\label{sec:formalism}
\vspace{-6pt}

\subsection{Model}

In this work, we consider matter composed of the full baryon octet, the~quartet of $\Delta$-resonances and leptons---electrons and muons
($e^-,\mu^-$). The~strong interaction between (non-strange) baryons is mediated by
the $\sigma,\omega$ and~$\rho$-mesons. In~addition, the~hidden-strangeness mesons $\sigma^*$ (scalar) and $\phi$ (vector)
mediate the hyperon--hyperon interactions. The~total Lagrangian density
of the fermionic component of matter in presence of a magnetic field is
given by,
\begin{equation} \label{eqn.1}
  \mathcal{L}= \mathcal{L}_m + \mathcal{L}_f,
\end{equation}
where, $\mathcal{L}_m$ and $\mathcal{L}_f$ are the matter and the
gauge field contributions, respectively.

We take the matter part of the Lagrangian density
as \citep{2015PhRvC..92a5802C,2017PASA...34...65T},
\begin{equation} \label{eqn.2}
\begin{aligned}
\mathcal{L}_m = & \sum_{b} \bar{\psi}_b(i\gamma_{\mu} D^{\mu} - m_b + g_{\sigma b}\sigma + g_{\sigma^* b}\sigma^* - g_{\omega b}\gamma_{\mu}\omega^{\mu} - g_{\phi b}\gamma_{\mu}\phi^{\mu} - g_{\rho b}\gamma_{\mu} \boldsymbol{\tau}_b \cdot \boldsymbol{\rho}^{\mu}) \psi_b \\
    & + \sum_{d} \bar{\psi}_{d\nu}(i\gamma_{\mu} D^{\mu} - m_{d}
    + g_{\sigma d}\sigma - g_{\omega
      d}\gamma_{\mu}\omega^{\mu} - g_{\rho d}\gamma_{\mu} \boldsymbol{\tau}_{\Delta} \cdot \boldsymbol{\rho}^{\mu}) \psi_{d}^{\nu} \\
	& + \frac{1}{2}(\partial_{\mu}\sigma\partial^{\mu}\sigma - m_{\sigma}^2 \sigma^2) + \frac{1}{2}( \partial_{\mu}\sigma^*\partial^{\mu}\sigma^*- m_{\sigma^*}^2 \sigma^{*2}) - \frac{1}{4}\omega_{\mu\nu}\omega^{\mu\nu} + \frac{1}{2}m_{\omega}^2\omega_{\mu}\omega^{\mu} \\
	& - \frac{1}{4}\phi_{\mu\nu}\phi^{\mu\nu} + \frac{1}{2}m_{\phi}^2\phi_{\mu}\phi^{\mu} - \frac{1}{4}\boldsymbol{\rho}_{\mu\nu} \cdot \boldsymbol{\rho}^{\mu\nu} + \frac{1}{2}m_{\rho}^2\boldsymbol{\rho}_{\mu} \cdot \boldsymbol{\rho}^{\mu} \\
	& + \sum_l \bar{\psi}_l (i\gamma_{\mu} D^{\mu} - m_l)\psi_l
\end{aligned}
\end{equation}
where $D^{\mu}= \partial^{\mu} + ieQ A^{\mu}$ is the covariant
derivative, $A^{\mu}$ is the electromagnetic vector potential, $eQ$ is the
charge of the particle ($e$ being unit `+' charge), the~$b$-summation in the first line is over the nucleons and hyperons
(spin-1/2), $d$-summation in the second line is over the
$\Delta$-resonances (spin-3/2) and the $l$ summation in the last line is
over leptons. The~fields $\psi_b$, $\psi_l$ and~$\psi_{d}^{\nu}$
correspond to the Dirac baryons, leptons and~the Rarita--Schwinger
fields for $\Delta$-resonances. Their masses are, respectively, $m_b$,
$m_l$ and $m_{d}$. The~third and fourth lines in
Equation~(\ref{eqn.2}) contain the contributions from scalar meson fields
$\sigma$ and $\sigma^*$ with masses $m_{\sigma}$ and $m_{\sigma^*}$,
isoscalar-vector meson fields $\omega_{\mu}$ and $\phi_{\mu}$ with
masses $m_{\omega}$ and $m_{\phi}$ and, finally, the~isovector-vector
meson field $\boldsymbol{\rho}_{\mu}$ with mass $m_{\rho}$. The~coupling between the mesons and baryons is described by the
density-dependent couplings $g_{ib}$ and $g_{i d}$, whereby 
{$\boldsymbol{\tau}_i$} stands for the iso-spin operator. Finally, the~purely ``gauge'' mesonic contributions in Equation~(\ref{eqn.2}) contain
the tensors
\begin{equation} \label{eqn.3}
\begin{aligned}
\omega_{\mu \nu} & = \partial_{\mu}\omega_{\nu} - \partial_{\mu}\omega_{\nu} ,\\
\phi_{\mu \nu} & = \partial_{\nu}\phi_{\mu} - \partial_{\mu}\phi_{\nu} ,\\
\boldsymbol{\rho}_{\mu \nu} & = \partial_{\nu}
\boldsymbol{\rho}_{\mu} - \partial_{\mu}\boldsymbol{\rho}_{\nu}.
\end{aligned}
\end{equation}

The electro-magnetic field Lagrangian density in Equation~(\ref{eqn.1}) has
the standard form
\begin{equation}\label{eqn.8}
\mathcal{L}_f = -\frac{1}{16 \pi} F_{\mu \nu} F^{\mu \nu}
\end{equation}
with $F^{\mu \nu}$ being the electro-magnetic field tensor.  Below,~we
adopt the reference frame in which the four-vector potential has the
form $A^{\mu} \equiv (0,-yB,0,0)$, where $B$ is the magnitude of
the magnetic~field.

In the mean-field approximation, assuming that the system is infinite,
the meson fields acquire the ground-state expectation values,
\begin{equation}\label{eqn.5}
\begin{aligned}
	\sigma = \sum_{b} \frac{1}{m_{\sigma}^2} g_{\sigma b}n_{b}^s + \sum_{d} \frac{1}{m_{\sigma}^2} g_{\sigma d}n_{d}^s,
 \quad \sigma^{*} = \sum_{b} \frac{1}{m_{\sigma^{*}}^2} g_{\sigma^{*}
   b} n_{b}^s\quad \text{(isoscalar-scalar)},
\end{aligned}
\end{equation}
\begin{equation}\label{eqn.6}
\begin{aligned}
  \omega_{0}= \sum_{b} \frac{1}{m_{\omega}^2} g_{\omega b}n_{b} +\sum_{d} \frac{1}{m_{\omega}^2} g_{\omega d}n_{d},
  \quad \phi_{0}= \sum_{b} \frac{1}{m_{\phi}^2} g_{\phi b}n_{b} \quad \text{(isoscalar-vector)},
\end{aligned}
\end{equation}
\begin{equation}\label{eqn.7}
\begin{aligned}
  \rho_{03}= \sum_{b} \frac{1}{m_{\rho}^2} g_{\rho b}
  \boldsymbol{\tau}_{b3}n_{b} + \sum_{d} \frac{1}{m_{\rho}^2} g_{\rho d}
  \boldsymbol{\tau}_{d3}n_{d}\quad\text{(isovector-vector)},
\end{aligned}
\end{equation}
where the scalar and baryon (vector) number densities are defined
 for the baryon octet as
$n_{b}^s= \langle\bar{\psi}_b \psi_b\rangle$  and~$n_{b}=
\langle\bar{\psi}_b \gamma^0 \psi_b\rangle$, respectively. For~
  the $\Delta$-resonances, these are defined as $n_{d}^s=
  \langle\bar{\psi}_{d\nu} \psi^\nu_d\rangle$ and $n_{d}=
  \langle\bar{\psi}_{d\nu} \gamma^0 \psi^\nu_d\rangle$,
  respectively. The~effective (Dirac) baryon masses in the same
approximation are given by,
\begin{equation}\label{eqn.4}
m_{b}^* = m_b - g_{\sigma b}\sigma - g_{\sigma^*b}\sigma^*, \quad
m_{d}^* = m_d - g_{\sigma d}\sigma
\end{equation}

The scalar density, baryon number density and the kinetic energy
density of the uncharged baryon (denoted by index $u$) at zero
temperature are given by,
\begin{eqnarray}\label{eqn.9}
n^{s}_u & = &\frac{2J_u +1}{2 \pi^2} m^{*}_{u} \left[ p_{{F}_{u}} E_{F_u}  - m_{u}^{*^2} \ln \left( \frac{p_{{F}_u} + E_{F_u}}{m_{u}^{*}} \right) \right], \\
n_u & = & (2J_u +1) \frac{p_{{F}_{u}}^{3}}{6 \pi^2},\\
\label{eqn.10}
\varepsilon_u &= &\frac{2J_{u} + 1}{2 \pi^2} \left[ p_{{F}_u} E^{3}_{F_u} - \frac{m_{u}^{*^2}}{8} \left( p_{{F}_u} E_{F_u} + m^{*^2}_{u} \ln \left( \frac{p_{{F}_u} + E_{F_u}}{m_{u}^{*}} \right) \right) \right],
\end{eqnarray}
respectively, where, $J_u$, $p_{{F}_{u}}$, $m^{*}_u$, $E_{F_u}$ are
the spin, Fermi momentum, effective mass and Fermi energy of the
$u$${\text{th}}$-uncharged~baryon.

The same quantities for a charged baryon (denoted by index $c$) are given by
the following~expressions:
\begin{itemize} [leftmargin=1.8em,labelsep=4.5mm]
\item Spin-1/2 case:
\begin{eqnarray}\label{eqn.11}
n^{s}_c & =& \frac{e|Q|B}{2 \pi^2} m_{c}^{*} \sum_{\nu=0}^{\nu_{max}}(2-\delta_{\nu,0}) \ln \left( \frac{p_{{F}_c} + E_{F_c}}{\sqrt{m^{*^2}_c + 2\nu e|Q|B}} \right), \\
n_c & = &\frac{e|Q|B}{2 \pi^2} \sum_{\nu=0}^{\nu_{max}}(2-\delta_{\nu,0})
p_{{F}_{c}},\\
\varepsilon_c &=&  \frac{e|Q|B}{2 \pi^2} \sum_{\nu=0}^{\nu_{max}}(2-\delta_{\nu,0}) \left[ p_{{F}_c} E_{F_c} + \left( m^{*^2}_c + 2\nu e|Q|B \right) \ln \left( \frac{p_{{F}_c} + E_{F_c}}{\sqrt{m^{*^2}_c + 2\nu e|Q|B}} \right)  \right],
\end{eqnarray}

\item Spin-3/2 case:
\begingroup\makeatletter\def\f@size{9}\check@mathfonts
\def\maketag@@@#1{\hbox{\m@th\normalsize\normalfont#1}}%
\begin{eqnarray}\label{eqn.13}
n^{s}_c & = &\frac{e|Q|B}{2 \pi^2} m_{c}^{*} \sum_{\nu=0}^{\nu_{max}}(4-\delta_{\nu,1}-2\delta_{\nu,0}) \ln \left( \frac{p_{{F}_c} + E_{F_c}}{\sqrt{m^{*^2}_c + 2\nu e|Q|B}} \right), \\
n_c & = &\frac{e|Q|B}{2 \pi^2} \sum_{\nu=0}^{\nu_{max}}(4-\delta_{\nu,1}-2\delta_{\nu,0}) p_{{F}_{c}},\\
\label{eqn.14}
\varepsilon_c &=  &\frac{e|Q|B}{2 \pi^2}
\sum_{\nu=0}^{\nu_{max}}(4-\delta_{\nu,1}-2\delta_{\nu,0}) \left[
  p_{{F}_c} E_{F_c} + \left( m^{*^2}_c + 2\nu e|Q|B \right) \ln \left(
    \frac{p_{{F}_c} + E_{F_c}}{\sqrt{m^{*^2}_c
        + 2\nu e|Q|B}} \right)  \right], \nonumber\\
\end{eqnarray}
\endgroup

\end{itemize}
where, $p_{{F}_{c}}$, $m^{*}_c$, $E_{F_c}$ are the Fermi
momentum of the $\nu^{\text{th}}$-Landau level, effective mass and
Fermi energy of the $c^{\text{th}}$-charged baryon.  The~Landau levels
for spin-$1/2$, $3/2$ baryons are denoted by $\nu$, the~maximum value of
which is defined by,
\begin{equation}\label{eqn.15}
\begin{aligned}
\nu_{max} = \text{Int} \left( \frac{p_{{F}_{c}}}{2e|Q|B} \right).
\end{aligned}
\end{equation}

In the case of spin-$1/2$ particles, the~lowest Landau level has
degeneracy equal unity and all other levels have degeneracy equal 2
~\cite{2013NuPhA.898...43S}.  In~the case of spin-$3/2$ particles, the~degeneracy of the lowest (first) level is 2, for~the second level it is
3 and~is 4 in the remaining Landau levels~\cite{2013JPhG...40e5007D}.

For the case of leptons ($l \equiv e^-,\mu^-$), the~number density and
contribution to the kinetic energy density is given by,
\begin{equation}\label{eqn.16}
n_l = \frac{e|Q|B}{2 \pi^2} \sum_{\nu=0}^{\nu_{max}}(2-\delta_{\nu,0})
p_{{F}_l},
\end{equation}
\begin{equation}\label{eqn.17}
\begin{aligned}
\varepsilon_l = {} & \frac{e|Q|B}{2 \pi^2}
\sum_{\nu=0}^{\nu_{max}}(2-\delta_{\nu,0}) \left[ p_{{F}_l} E_{F_l} +
  \left( m^{2}_l + 2\nu e|Q|B \right) \ln \left( \frac{p_{{F_l}} +
      E_{F_l}}
    {\sqrt{m^{2}_l + 2\nu e|Q|B}} \right)  \right],
\end{aligned}
\end{equation}
where, $p_{{F}_l}$, $m_l$, $E_{F_l}$ are the Fermi momentum of the
$\nu$${\text{th}}$-Landau level, bare mass and Fermi energy of the
lepton,~respectively.

In Equations~(\ref{eqn.9})--(\ref{eqn.14}), the~Fermi momenta $p_{F_u}$ and
$p_{F_c}$ are defined as,
\begin{equation}\label{eqn.18}
\begin{aligned}
p_{F_u} = \sqrt{E^{2}_F - m^{*^2}}, \quad
p_{F_c} = \sqrt{E^{2}_F - (m^{*^2}  + 2\nu e|Q|B)},
\end{aligned}
\end{equation}
with $E_F$ being the Fermi energy of the respective particle. The~total energy density of the matter is thus  given by,
\begin{equation}\label{eqn.19}
\begin{aligned}
\varepsilon_m = \sum_{b} \varepsilon_b + \sum_{d} \varepsilon_d + \frac{1}{2}m_{\sigma}^2 \sigma^{2} + \frac{1}{2}m_{\sigma^*}^2 \sigma^{*2} +\frac{1}{2} m_{\omega}^2 \omega_{0}^2 + \frac{1}{2} m_{\phi}^2 \phi_{0}^2 + \frac{1}{2}m_{\rho}^2 \rho_{03}^2 + \sum_l \varepsilon_l
\end{aligned}
\end{equation}
where the~sum over $b,d$
includes the 
baryon octet and the $\D$-quartet,
and the $l$-summation is over the leptons. The~matter pressure can be evaluated from the
thermodynamic (Gibbs--Duhem) relation as,
\begin{equation}\label{eqn.20}
p_m = \sum_{b} \mu_b n_b + \sum_{d} \mu_d n_d + \sum_{l} \mu_l n_l - \varepsilon_m,
\end{equation}
where $\mu_{b(d)}=\partial \varepsilon_m/ \partial n_{b(d)}$ is the chemical potential of the $b{\text{th}}$-spin-1/2 ($d{\text{th}}$-spin-3/2) baryon which can be defined as,
\begin{equation}\label{eqn.21}
\begin{aligned}
& \mu_{b} = \sqrt{p_{F_b}^2 + m_{b}^{*2}} + g_{\omega b}\omega_{0} + g_{\phi b}\phi_{0} + g_{\rho b} \boldsymbol{\tau}_{b3} \rho_{03} + \Sigma^{r}, \\
& \mu_{d} = \sqrt{p_{F_d}^2 + m_{d}^{*2}} + g_{\omega d}\omega_{0} +  g_{\rho d} \boldsymbol{\tau}_{d3} \rho_{03} + \Sigma^{r}
\end{aligned}
\end{equation}

In order to maintain thermodynamic consistency a self-energy re-arrangement term, $\Sigma^{r}$~is~introduced
\begin{equation}\label{eqn.22}
\begin{aligned}
\Sigma^{r} & = \sum_{b} \left[ \frac{\partial g_{\omega b}}{\partial n}\omega_{0}n_{b} - \frac{\partial g_{\sigma b}}{\partial n} \sigma n_{b}^s + \frac{\partial g_{\rho b}}{\partial n} \rho_{03} \boldsymbol{\tau}_{b3} n_{b} - \frac{\partial g_{\sigma^{*} b}}{\partial n} \sigma^{*} n_{b}^{s} + \frac{\partial g_{\phi b}}{\partial n}\phi_{0}n_{b} \right] \\
& +\sum_d \left[ \frac{\partial g_{\omega d}}{\partial n}\omega_{0}n_{d} - \frac{\partial g_{\sigma d}}{\partial n} \sigma n_{d}^s + \frac{\partial g_{\rho d}}{\partial n} \rho_{03} \boldsymbol{\tau}_{d3} n_{d} \right]
\end{aligned}
\end{equation}
where $n= \sum_{b} n_b + \sum_d n_d$ is the total baryon (or vector) number
density. This re-arrangement term contributes  explicitly only to the matter
pressure.  {Equations (\ref{eqn.19}) and (\ref{eqn.20}) provide the
  EoS (the relation between the pressure and energy density) of the
  system under consideration in a
  parametric form. 
  }

\subsection{Meson--Baryon Coupling~Parameters}

In this work, we adopt the DD-ME2 %please define if appropriate. 
 parameterization proposed in
Ref.~\cite{2005PhRvC..71b4312L} for nucleonic matter. The~density-dependent meson--nucleon coupling constants are given by,
\begin{equation}\label{eqn.23}
g_{i N}(n)= g_{i N}(n_{0}) f_i(x) \ \ \ \ \ \ \text{for }i=\sigma,\omega,
\end{equation}
where $x=n/n_0$, $n_0$ is the saturation density, $N$ stands for a
nucleon and~\begin{equation}\label{eqn.24}
f_i(x)= a_i \frac{1+b_i (x+d_i)^2}{1+c_i (x+d_i)^2}.
\end{equation}

For the $\rho$-meson, the~density-dependent coupling constant is given by
\begin{equation}\label{eqn.25}
g_{\rho N}(n)= g_{\rho N}(n_{0}) e^{-a_{\rho}(x-1)}.
\end{equation}

The details of the procedure for fixing the values of coefficients in
Equations~(\ref{eqn.24}) and (\ref{eqn.25}) can be found in
Ref.~\cite{2005PhRvC..71b4312L}. Table~\ref{tab:1} provides the
parameter values employed in this work. Note that nucleons do not
couple with the $\sigma^*$ and $\phi$-mesons, i.e.,
$g_{\sigma^* N}= g_{\phi N}=0$.

\begin{table}[H]
\centering
\caption{The coupling constants for the DD-ME2 %please define if appropriate. 
  parameterization~\citep{2005PhRvC..71b4312L},
  at nuclear saturation density $n_0=0.152$ fm$^{-3}$. For~this model,
  the nuclear parameters are:
  the compression modulus $K_0=250.89$~MeV, the~binding energy per
  nucleon $E/A=-16.14$ MeV,
  the symmetry energy $a_{sym}=32.3$~MeV, the~effective nucleon Dirac
  mass $m^*_N/m_N =0.572$ with $m_N=938.9$ MeV.}
\begin{tabular}{ccccccc}
\toprule
\textbf{Meson} \textbf{(\boldmath{$i$})} & \boldmath{$m_i$}\textbf{ (MeV) }& \boldmath{$g_{iN}(n_0)$} & \boldmath{$a_i$} & \boldmath{$b_i$} & \boldmath{$c_i$} & \boldmath{$d_i$} \\
\midrule
$\sigma$ & 550.1238 & 10.5396 & 1.3881 & 1.0943 & 1.7057 & 0.4421 \\
$\omega$ & 783 & 13.0189 & 1.3892 & 0.9240 & 1.4620 & 0.4775 \\
$\rho$ & 763 & 7.3672 & 0.5647 & & & \\
\bottomrule
\end{tabular}
\label{tab:1}
\end{table}

For the hyperonic sector, the~density-dependent vector coupling
constants are determined from SU(6) symmetry~\cite{2001PhRvC..64e5805B}
\begin{equation}\label{eqn.26}
\begin{aligned}
\frac{1}{2}g_{\omega \Lambda} & = \frac{1}{2}g_{\omega \Sigma}=g_{\omega \Xi}= \frac{1}{3}g_{\omega N}, \\
2g_{\phi \Lambda} & =2g_{\phi \Sigma}=g_{\phi \Xi}= -\frac{2\sqrt{2}}{3}g_{\omega N}, \\
\frac{1}{2}g_{\rho \Sigma} & =g_{\rho \Xi}=g_{\rho N}, \ \ g_{\rho \Lambda}=0.
\end{aligned}
\end{equation}

For the evaluation of hyperon--$\sigma$-meson coupling constants, we
consider the optical potentials of the $\Lambda, \Sigma$ and $\Xi$
hyperons in nuclear matter to be, $U^{(N)}_{\Lambda}(n_0)=-30$ MeV,
$U^{(N)}_{\Sigma}(n_0)=+30$ MeV and $U^{(N)}_{\Xi}(n_0)=-14$ MeV. Due
to the repulsive nature of the $\Sigma$-potential in nuclear matter,
$\Sigma$-hyperons do not matter for the densities considered in
this work. 
 The~$\sigma^*$--$\Lambda$ coupling constant is evaluated by
fitting it to a potential depth
$U^{(\Lambda)}_{\Lambda}(n_0/5)=-0.67$ MeV and further constraining
the $\sigma^*$--$\Xi$ and $\sigma^*$--$\Sigma$ couplings via the
relation~\cite{2018EPJA...54..133L}
\begin{equation}\label{eqn.27}
\frac{g_{\sigma^* Y}}{g_{\phi Y}}=\frac{g_{\sigma^* \Lambda}}{g_{\phi
    \Lambda}},
\ \ Y \in \text{$\{\Xi,\Sigma\}$}.
\end{equation}

Table~\ref{tab:2} provides the numerical values for the density-dependent scalar couplings.
The coupling constants for the $\Delta$-resonances are taken as~\cite{2019PhRvC.100a5809L}
\begin{equation}\label{eqn.28}
g_{\omega d}= 1.1 g_{\omega N}, \ \ \ g_{\rho d}= g_{\rho N}.
\end{equation}

The density-dependent $g_{\sigma d}$ scalar coupling is determined by
fixing the $\Delta$-potential to the value $V_{\D} = \frac{4}{3}V_N$,
where $V_N$ is the isoscalar nucleon potential.  This implies that
$g_{\sigma d}/g_{\sigma N} =1.16$. Note that $\Delta$-resonances do not
couple to $\sigma^*$ and $\phi$-mesons, i.e.,
$g_{\sigma^* d}=g_{\phi d}=0$.
\begin{table}[H]
\centering
\caption{Scalar meson--hyperon coupling~constants.}
\begin{tabular}{cccc}
\toprule
 & \boldmath{$\Lambda$} & \boldmath{$\Xi$} & \boldmath{$\Sigma$} \\
\midrule
$g_{\sigma Y}/g_{\sigma N}$ & 0.6105 & 0.3024 & 0.4426 \\
$g_{\sigma^* Y}/g_{\sigma N}$ & 0.4777 & 0.9554 & 0.4777 \\
\bottomrule
\end{tabular}
\label{tab:2}
\end{table}
\unskip
\subsection{Magnetic Field~Profiles}

To model the magnetic field profile in the neutron star
interior, we adopted two types of profiles which give the dependence
of the field on the position inside the~star.

The {\it{exponential profile}} %italics removed. please confirm. 
is given by~\cite{1997PhRvL..79.2176B}
\begin{equation}\label{eqn.29}
B \left( \frac{n_b}{n_0} \right) = B_s + B_c \left\lbrace 1 - \exp
  \left[ -\beta \left( \frac{n_b}{n_0} \right)^{\gamma} \right]
\right\rbrace ,
\end{equation}
where $B_s$ and $B_c$ denote the magnetic fields at surface and
at center of the star, respectively. The~free
parameters $\beta$ and $\gamma$ are commonly adjusted such as to
have a fixed surface magnetic field of $10^{15}$ G for any given value
of the field in the center, which is typically larger than the surface~field.

The {\it{universal profile}%Is the italics necessary?
} is given as~\cite{2019PhRvC..99e5811C}
\begin{equation}\label{eqn.30}
B(x)= B_c \left(1 - 1.6x^2 - x^4 + 4.2x^6 -2.4x^8\right),
\end{equation}
where $x=r/r_{\rm mean}$, $r$ is the internal radius joining the
center to the  point of observation, $r_{\rm mean}$ is the mean radius of the
star and $B_c$ is the value of the field at the center of the~star.

\section{Results}
\label{sec:results}

We turn now to the numerical results of our study and consider
the  effect of strong magnetic field on high-density stellar matter with three types
of~composition:
\begin{itemize}[leftmargin=2.2em,labelsep=4mm] %[noitemsep,topsep=0.5pt]
\item[(1)] Nucleons (N),
\item[(2)] nucleons and hyperons (NY),
\item[(3)] nucleons, hyperons and $\D$-resonances (NY$\Delta$).
\end{itemize}

\begin{figure}[H]
  \centering
\includegraphics[width=12.5cm,keepaspectratio ]{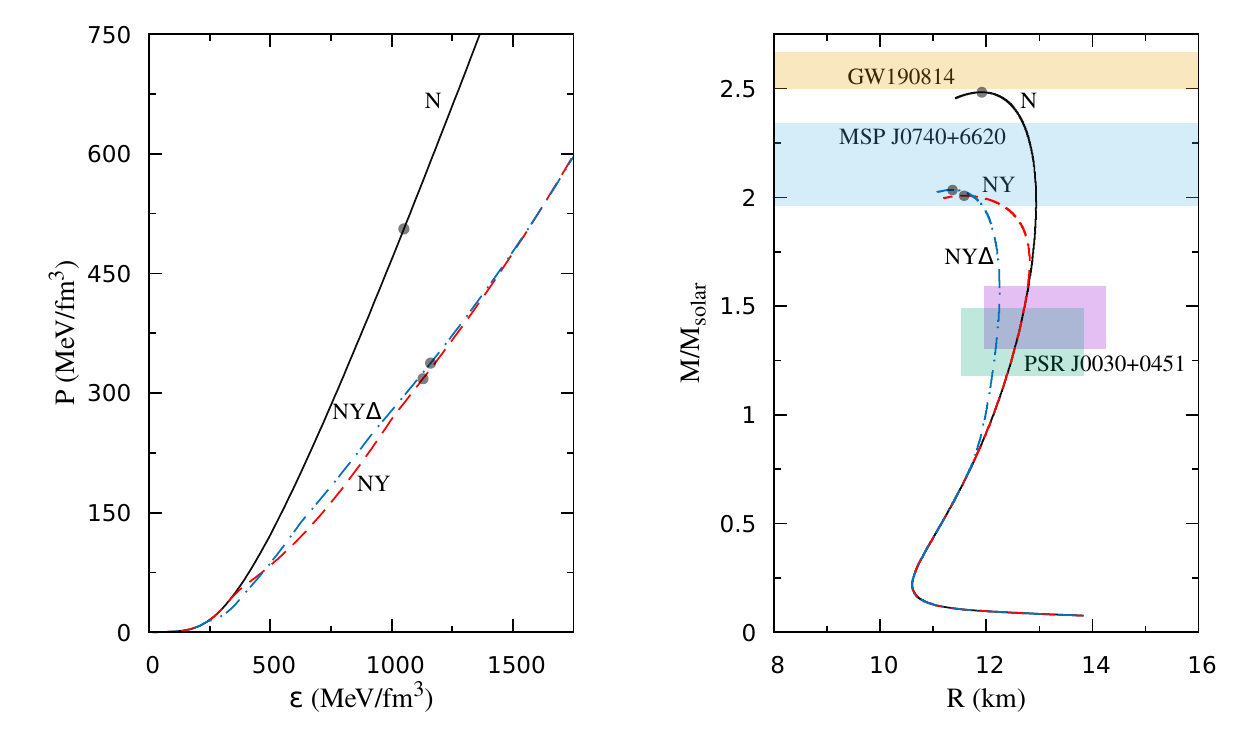}
\caption{Left panel: Equation of state (EoS) of matter in the absence of magnetic field
  for three compositions considered.  Solid, dashed, dash-dotted lines
  correspond to the cases of nucleons only (N), nucleons and hyperons
  (NY), nucleons, hyperons and Delta (NY$\Delta$), respectively.  The circles indicate the location of the maximum mass star for each
  composition.  Right panel: the mass--radius ($M$--$R$) relations corresponding to
  the EoS on the left panel obtain through solutions of the TOV %please define if appropriate. 
  equations. The~mass constraints from the various astrophysical
  observations are represented by the colored bands and correspond to
  the GW190814 event~\cite{2020ApJ...896L..44A}, MSP %Please define if appropriate. 
   J0740+6620
~\cite{2020NatAs...4...72C} and the mass--radius limits inferred for
  PSR %please define if appropriate. 
  J0030+0451 from the Neutron star Interior Composition ExploreR (NICER) experiment
~\cite{2019ApJ...887L..24M,2019ApJ...887L..21R}.}
\label{fig-1}
%\end{center}
\end{figure}

Figure~\ref{fig-1} shows the EoS and mass--radius (hereafter $M$--$R$)
relations for these three compositions of matter in the case $B=0$
along with some astrophysical constraints on the masses and radii of
compact stars.  Table~\ref{table-3} lists some parameters of the stars
with maximum masses along the stellar sequences for the three
compositions listed~above.

Let us now turn to stellar configurations with magnetic field.  In~order to show the effect of the field in the case of the {\it{universal profile}}, we need a relation between the internal radius
and the density, i.e.,~the function $r(n)$. This requires us to
specify a stellar model. We, therefore, chose three stars from the
stable region of $M$--$R$ curve with parameters shown in
Table~\ref{table-4}.  After~fixing the value of the central field
$B_c=2.9\times10^{18}$ G, we are then able to use Equation~(\ref{eqn.30}).
We note that the predicted surface magnetic field values are
$B_s\approx 5.6\times 10^{17}$ G.

In the case of {exponential profile} (\ref{eqn.29}), the surface field is
fixed at $B_s=10^{15}$ G and we adopt the parameter values
 $\beta=0.01$ and $\gamma=3.95,~3.15$ and $3.2$ for N, NY and~NY$\Delta$ matter, respectively.  The~resulting magnetic field profiles
guarantee that the matter remains stable under the influence of strong~$B$-field.

\unskip

\begin{table}[H]
\centering
\caption{Parameter values of the maximum-mass stars for non-magnetic
  stellar sequences derived for three different compositions. Here,
  $M_{max}$, $R$, $\varepsilon_c$ denote the maximum mass (in solar
  units), corresponding radius (in km) and central energy density (in
  MeV/fm$^3$), respectively.}\label{table-3}
\begin{tabular}{cccc}
\toprule
\textbf{Composition} & \boldmath{$M_{max}$} \textbf{(\boldmath{$M_\odot$})} & \boldmath{$R$} \textbf{(km)} & \boldmath{$\varepsilon_c$} \textbf{(MeV/fm$^3$)} \\
\midrule
N & 2.483 & 11.941 & 1035.558 \\
NY & 2.008 & 11.606 & 1119.700 \\
NY$\Delta$ & 2.034 & 11.365 & 1161.771 \\
\bottomrule
\end{tabular}
\end{table}
\unskip
\begin{table}[H]
\centering
%\caption{Same as in Table~\ref{table-3}, but for magnetic stars with
%   profiles.}\label{table-4}
\caption{The values of the energy--density, pressure and number density
  at the center of {{non-magnetized}%Is the italics necessary?
} stellar models which were used
  to obtain the universal relation (\ref{eqn.30}) as a function of
  density (instead of the internal radius). We also list the mass and
  the radius for each model.}\label{table-4}
\begin{tabular}{cccccc}
\toprule
\textbf{Composition} & \textbf{\boldmath{$M$}(\boldmath{$M_\odot$})} & \textbf{$R$(km)} & \boldmath{$\varepsilon_c$}\textbf{(MeV/fm$^3$)} & \boldmath{$p_c$}\textbf{(MeV/fm$^3$)} & \boldmath{$n_c$}\textbf{(fm$^{-3}$)} \\
\midrule
N & 2.482 & 12.002 & 1000 & 467.44 & 0.788 \\
NY & 2.000 & 11.801 & 1000 & 266.94 & 0.846 \\
NY$\Delta$ & 2.034 & 11.376 & 1150 & 333.38 & 0.944 \\
\bottomrule
\end{tabular}
\end{table}

\begin{figure}[H]
  \centering
\includegraphics[width=11cm,keepaspectratio ]{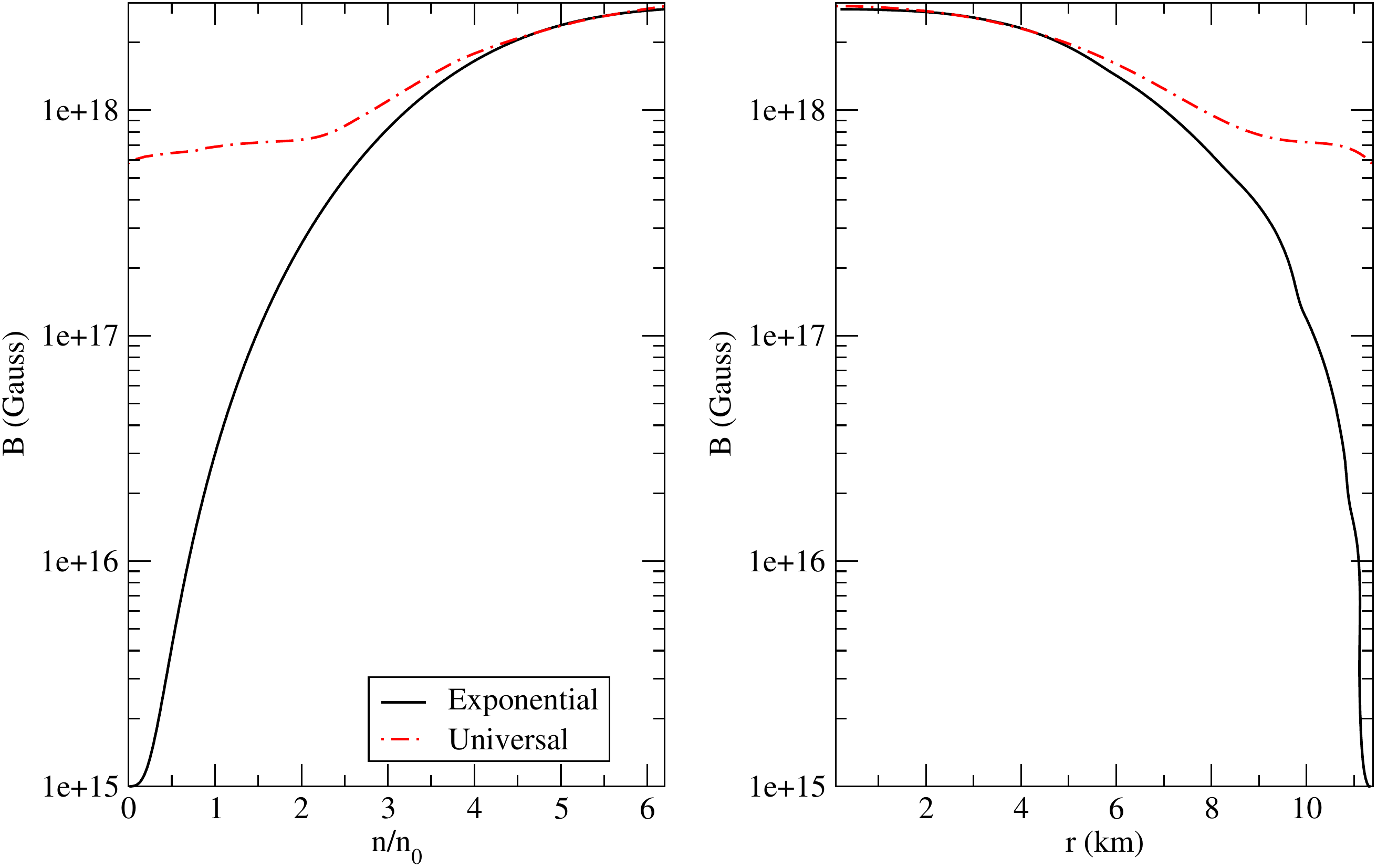}
\caption{Magnetic field profiles for the NY$\Delta$ composition as a
  function of baryon number density, $n$~(\textbf{left panel}) and internal
  radius $r$ (\textbf{right panel}). Black solid, red dash-dotted lines denotes
  exponential and universal profiles, respectively.}
\label{fig-2}
%\end{center}
\end{figure}

Figure~\ref{fig-2} shows the variationof the magnetic field in the
interior of the star as a function of internal radius and density in the cases
of exponential and universal~relations. %Please use scientific notation to represent the Ordinate value, for example, 1e+15 should revised as 1\times 10^{15}

\begin{figure}[H]
  \centering
\includegraphics[width=12cm,keepaspectratio ]{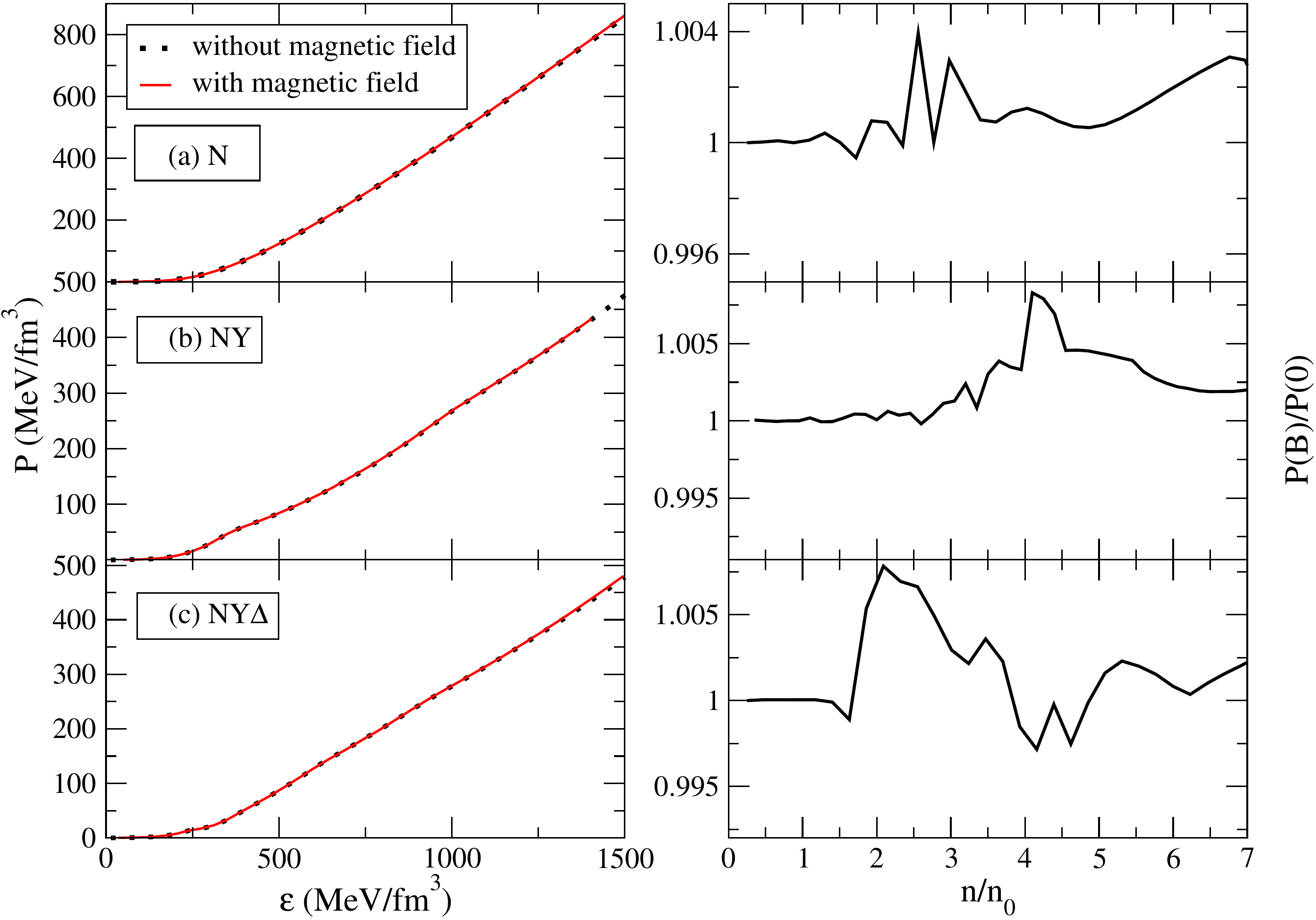}
\caption{A comparison of the EoS of magnetized and non-magnetized
  matter for three different compositions (\textbf{a}) nucleons only (N), (\textbf{b})
  nucleons and hyperons (NY) and~(\textbf{c}) nucleons, hyperons and
  $\Delta$-resonances. We assume  an exponential
  magnetic field profile. Left panels show the dependence of pressure
  on energy density with (solid lines) and without (dots) magnetic
  field. The~right panels the ratio of pressure in the
  presence of magnetic field $P(B)$  to that in the absence of the
  field $P(0)$ as a function of particle number density normalized to
  the nuclear saturation density. 
}
\label{fig-3}
%\end{center}
\end{figure}

\begin{figure}[H]
\centering
\includegraphics[width=12cm,keepaspectratio ]{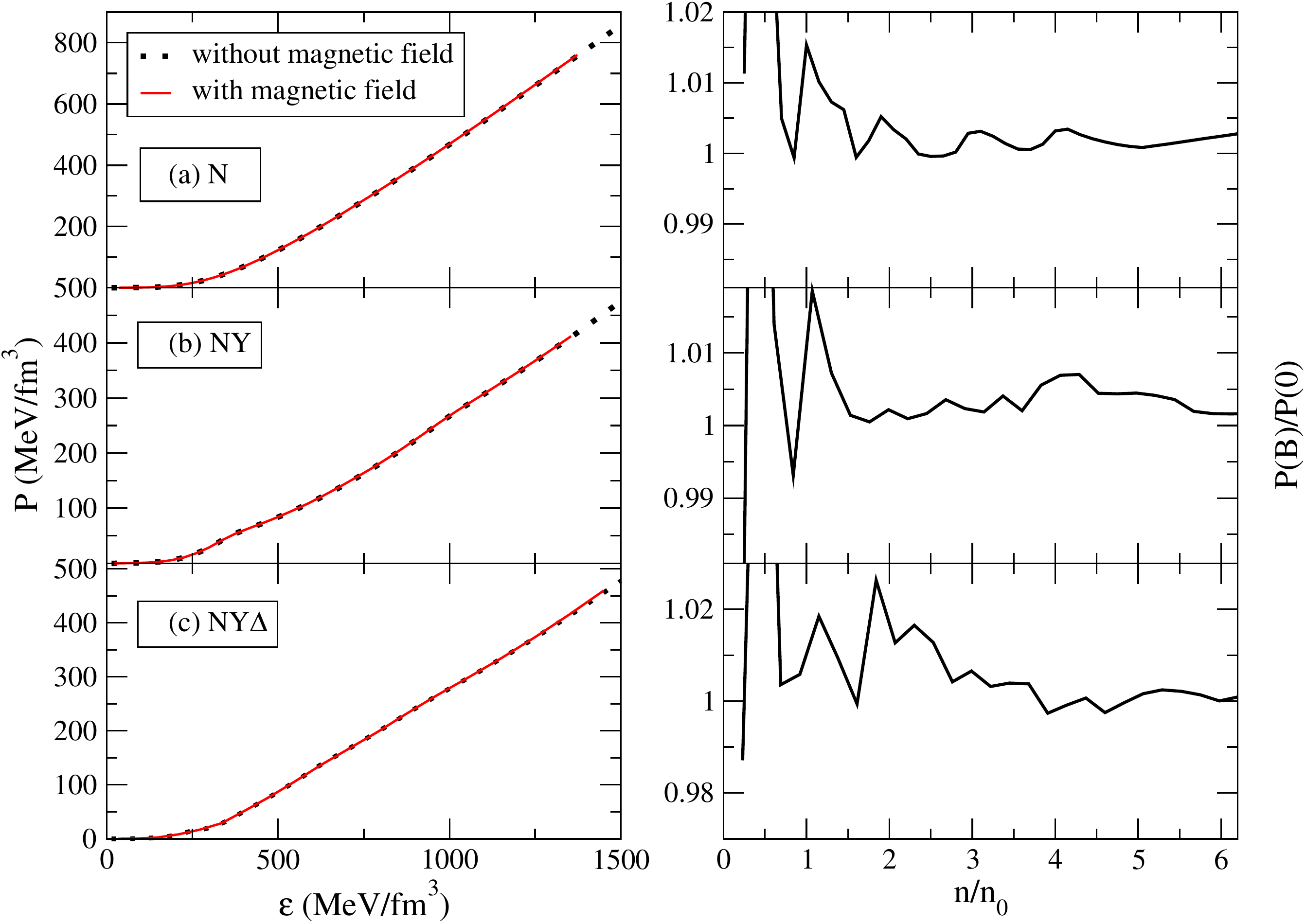}
\caption{Same as in Figure~\ref{fig-3}, but~for the universal
  magnetic field profile.}
\label{fig-4}
%\end{center}
\end{figure}

Figures~\ref{fig-3} and \ref{fig-4} show the EoS in the presence of
magnetic field for various compositions. In~all the cases studied, it is
seen that the EoSs follow the same trends with and without magnetic
field. For~the assumed values of the field, the changes in the EoS are
marginal if viewed on the $P(\varepsilon)$ plots shown on the left
panels of Figures~\ref{fig-3} and \ref{fig-4}.

\unskip

The right panels of the same figures show the ratio of the pressure in
presence of magnetic field to the pressure in the absence of the
field. The~oscillations in the pressure are associated with the occupation
of the Landau levels in the strong field. It is seen that these
oscillations are stronger at the surface of the star for the universal
profile, because~the field does not decay in this case as quickly as
for the exponential profile. In~the high density regime, the~oscillations are comparable for both the profiles. This behavior is a
consequence of the fact that close to the centre of the star, both
profiles have similar values of the magnetic field (see
Figure~\ref{fig-2}).

\begin{figure}[H]
 \centering
\includegraphics[width=12cm,keepaspectratio ]{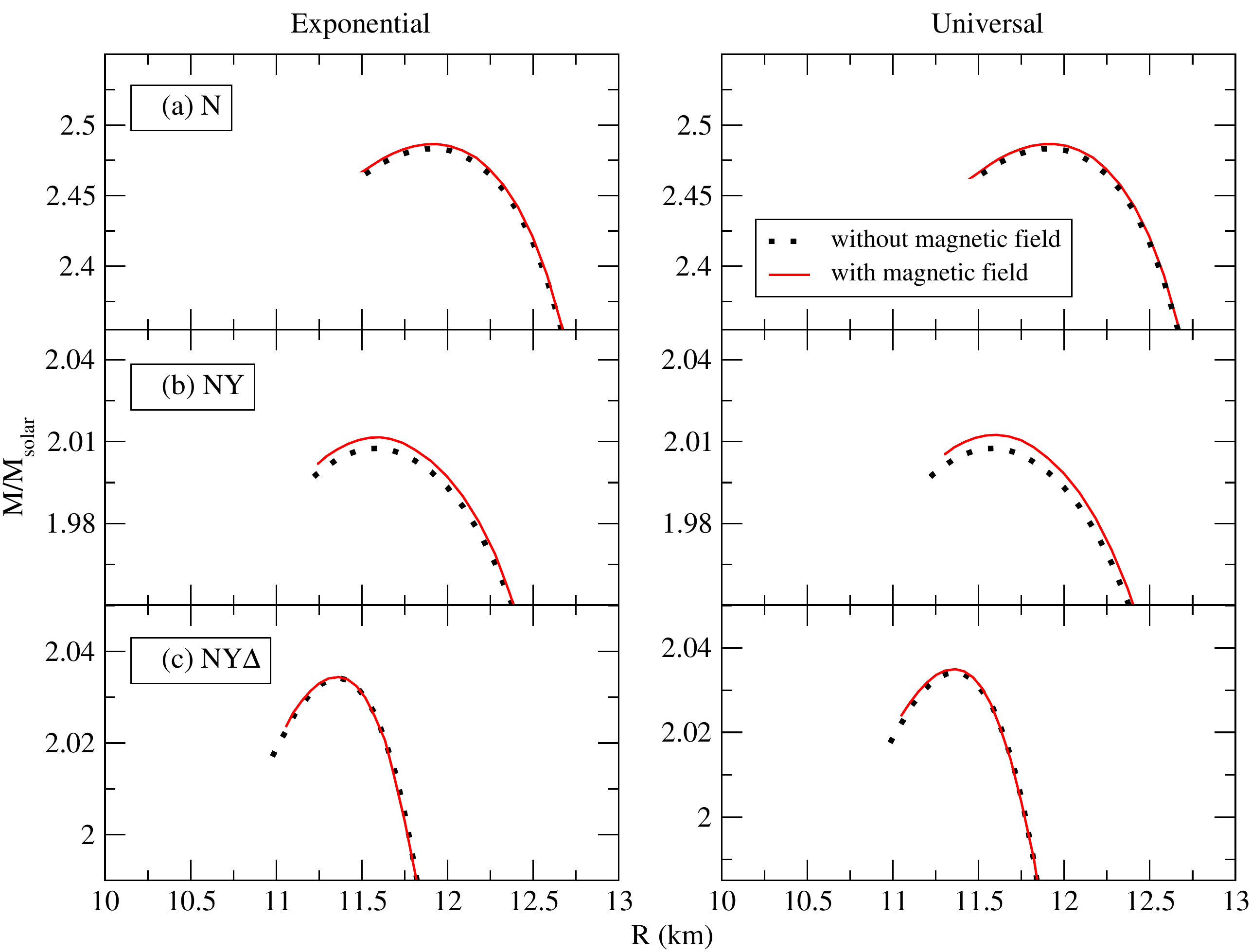}
\caption{The $M$--$R$ relations for three compositions considered: (\textbf{a}) nucleons
  only (N), (\textbf{b}) nucleons and hyperons (NY) and~(\textbf{c}) nucleons, hyperons
  and $\Delta$-resonances in the cases of with (solid lines) and
  without (dots) magnetic field. The~left panels correspond to the
  exponential magnetic field profile and the right panels to
  universal magnetic field profile.}
\label{fig-5}
%\end{center}
\end{figure}

As a result of the additional pressure due to the magnetic field, the~EoS
is stiffened and, consequently, the~maximum masses of magnetized
compact stars are higher compared to their non-magnetized
counterparts. This can be seen from Figure~\ref{fig-5}, where the
corresponding $M$--$R$ relations are displayed. More quantitatively, we
find that for the N-composition the increase in maximum mass due to
the effect of the magnetic field is about $0.13\%$, in~the case of
NY-composition $0.20\%$ for the exponential profile and $0.244\%$ for
the universal profile and, finally, for~the NY$\Delta$-composition,
about $0.01\%$ for the exponential profile and $0.034\%$ for the
universal profile. We note that in the case of NY-composition the EoS
is softer at high densities, than~in the case of
NY$\Delta$-composition. Therefore, the~effect of the magnetic field is
more sizable in the case of the softer EoS, {i.e.}, for~the
NY-composition. Thus, we conclude that the changes in the maximum mass
are more pronounced in the case of NY-composition and are less
significant for the NY$\Delta$-composition.

\begin{figure}[H]
  \centering
\includegraphics[width=11.5cm,keepaspectratio ]{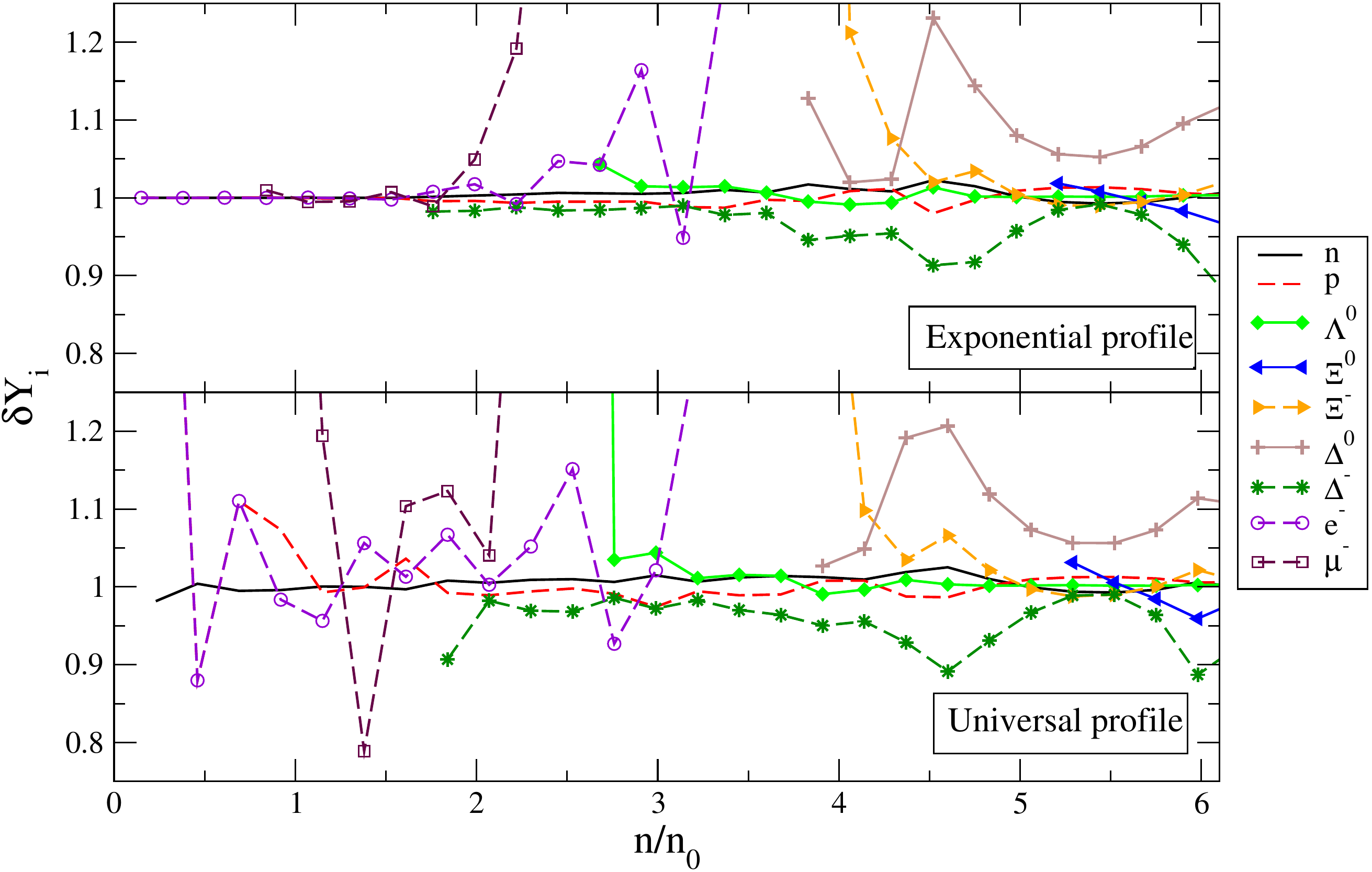}
\caption{ Dependence of the ratio $\delta Y_i = n_i(B)/n_i(0)$ on the
  baryon number density normalized to the nuclear saturation density
  $n_0$ for neutrons $(n)$, protons $(p)$, $\Lambda^0$,
  $\Xi^0$, $\Xi^-$, $\Delta^0$, $\Delta^-$, $e^-$ and $\mu^-$.
  The particle markers are indicated in the panel on the right. The  upper panel corresponds to the exponential field profile, the~lower one to the universal field profile.
}
\label{fig-6}
%$\end{center}
\end{figure}

Figure~\ref{fig-6} shows the ratio of fractions of different species
$\delta Y_i = n_i(B)/n_i(0)$ as a function of normalized baryon number
density. The~oscillating nature of the fractions arises due to
successive occupation of Landau levels for the charged species. The~effect of the field is not substantial in the low-density regime for
exponential field profile as the field strength in this case is small
near the surface. In~the case of the universal profile, the~low-density regime shows strong fluctuations because the decay of the
magnetic field with density is small and the surface magnetic field is
of the order few times of $10^{17}$ G (see Figure~\ref{fig-2}). It
  is interesting to note that for most of the particles $\delta Y_i >
  1$, but~in the case of $\Delta^-$, the~opposite is the case.  This
  is a consequence of the charge neutrality. Due to the Landau quantization
  the fraction of electrons increases compared to non-magnetic case
  which causes the $\Delta^-$ fraction to decrease.  The~pattern in
Figure~\ref{fig-6} results from the complicated interplay between the
Landau quantization for charge particles entering into the two key
conditions imposed---the charge neutrality and baryon number
conservation, which are used in the construction of the EoS. Note also that
  $\Delta^+$ and $\Delta^{++}$ resonances do not appear until baryon density of $n
  \geq 6.1 n_0$ for our particular choice of $\Delta$-potential.

\begin{figure}[H]
  \centering
\includegraphics[width=9cm,keepaspectratio ]{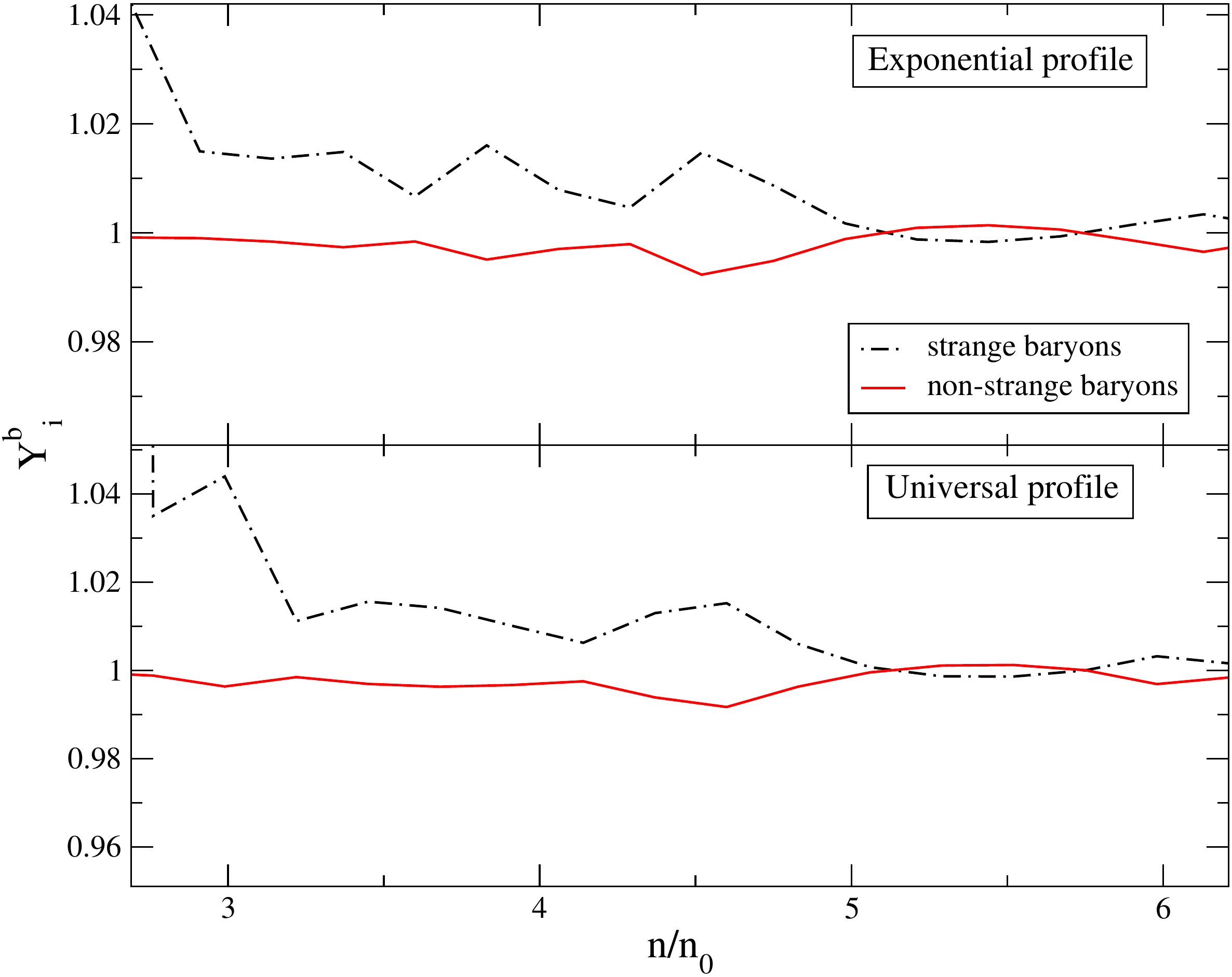}
\caption{
The ratio $Y_i^{b}$  of the partial fractions of strange or non-strange
  baryons in presence of a magnetic field to that without a magnetic
  field as a function of baryon number density, $n$ in units of $n_0$
  in the case of $NY\D$ composition.  Upper
  panel corresponds to the exponential field profile, the~lower panel to the universal field
  profile.}
\label{fig-7}
%$\end{center}
\end{figure}

  In Figure~\ref{fig-7} we show the quantity $Y_i^{b}$, which is defined
  as the ratio of the partial fractions of strange or non-strange
  baryons in presence of a magnetic field to that without a magnetic
  field. The~fraction of strange baryons is affected significantly
  ($\sim$4$\%$) by the magnetic field, whereas the fraction of the non-strange
  baryons is affected much less.  We see that in
  the presence of magnetic field strange baryons appear earlier than
  in the non-magnetic case. This is, again, a~consequence of complex interplay
  between the imposed charge neutrality condition and modifications of
  the phase-space occulation due to the Landau~quantization.

\begin{figure}[H]
   \centering
\includegraphics[width=9cm,keepaspectratio ]{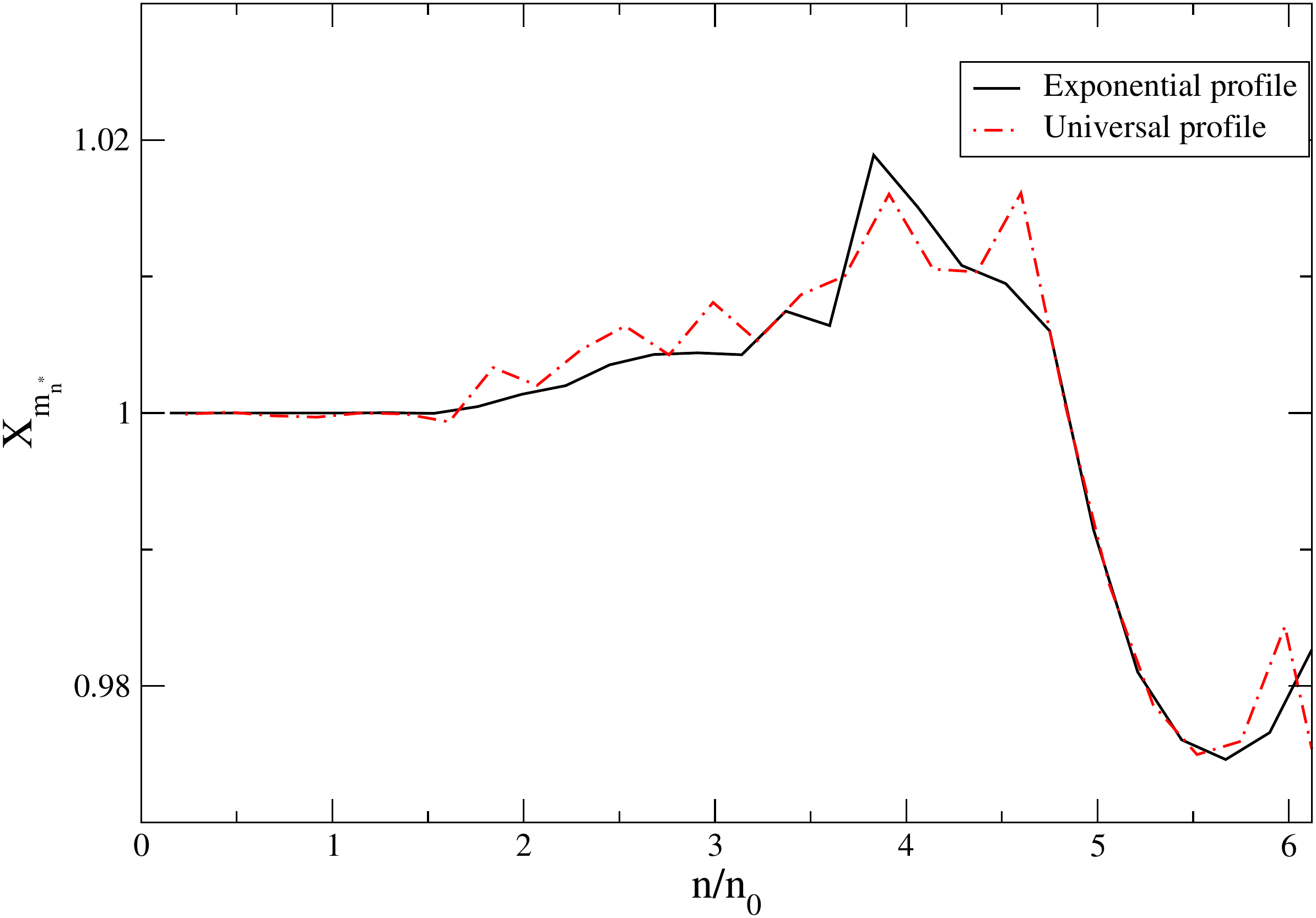}
\caption{
  Dependence of the ratio $X_{m_n^*} = m_N^*(B)/m_N^*(0)$
  of the effective nucleon Dirac mass in the magnetic field
to its value in the absence of the field
on baryon number density $n$ in units of $n_0$
  in the case of $NY\D$ composition. The~result  for the exponential
  field profile is shown by a solid  and for 
  universal by a dash-dotted line.}
\label{fig-8}
%\end{center}
\end{figure}

Finally, to~quantify the variations of the effective mass of a baryon
in the presence of magnetic field, we show in Figure~\ref{fig-8} the
ratio of effective nucleon Dirac mass ($m_N^*$) in the magnetic field
to its value in the absence of the field
$X_{m_n^*} = m_N^*(B)/m_N^*(0)$.  It is seen that, for~the exponential
profile case, $m_N^*$ remains unchanged until the appearance of
$\Delta^-$ around $1.3$ times nuclear saturation density.  A~similar
trend is observed for the universal profile case, but~the amplitudes
of the oscillations are larger. This is (again) due to the fact that the
magnetic field value at the surface of the star is larger for this profile.
With the onset of $\Xi^-$, we observe a reduction in $X_{m_n^*}$ by
about $4\%$ in the density range $\sim$4--5 times saturation density
for both the~profiles.

\unskip
\unskip

\unskip

\section{Conclusions and~Outlook}
\label{sec:conclusions}

Recent years have seen substantial progress in describing compact
stars with heavy baryons (hyperons as well as $\D$-resonances) in a
manner consistent with all the currently available astrophysical as well
as laboratory data. Motivated by this, we have extended, for~the first
time, one of the standard approaches which is based on CDF theory with
density-dependent couplings to the case of strongly magnetized
matter. In~doing so, we have taken into account fully the Landau
quantization of orbits of charged particles in strong fields.  We
confirm previous findings that magnetic fields make EoS stiffer and
lead to higher maximum masses of compact stars. To~quantify these
effects we employed two parameterizations of the magnetic field
profiles, namely the exponential~\cite{1997PhRvL..78.2898C} and the
universal~\cite{2019PhRvC..99e5811C} profiles  for a fixed value of
the  central
magnetic field $B_c= 2.9 \times 10^{18}$ G. The~universal
profile implies a relatively high surface magnetic field of
$\sim$5.6 $\times$ $10^{17}$ G and flat magnetic field profile. The~exponential profile, by~construction, is adjusted to produce a surface
magnetic field value $10^{15}$ G. In~this case, the profile is steep
with the magnetic field changing by orders of magnitude. Having the
profiles at hand, we~have explored the changes in the composition of
matter and the effective mass of the nucleon. We~find typical for magnetized
system oscillations in these quantities which are similar to the de
Haas-van Alphen oscillations of observables (e.g.,  the magnetic
susceptibility of electronic systems) in magnetic fields. The~oscillations have their origin in the occupation of the Landau levels
by particles. As a result that~the charged and neutral baryons are
coupled by the baryon number and charge conservation, the~oscillations
are coupled as well and affect the fractions of neutral particles
(neutrons, $\Lambda$s and $\Delta^0$s). The~oscillations of the
particle fractions are substantially different for the two profiles
studied if they are compared for the same value of the central magnetic
field. In~the case of the universal profile, these~oscillations extend up
to the low-density regime because the field does not change
substantially. In~the exponential profile case, the~low-density regime has low
magnetic fields, therefore the amplitudes of oscillations
are low. Comparing the oscillations in the strange and non-strange
sectors we observe that the hyperon fractions are more affected by the
magnetic fields that the non-strange baryon fractions within the
density range considered. Furthermore, the~Dirac nucleon effective
masses exhibit similar oscillations, which implies that a range of
quantities (specific heat, baryon mean-free-path, thermal~conductivity, etc.) may show oscillations as~well.

Our extension of the CDF-based EoS to the non-zero magnetic field can be
used to study a range of phenomena in and with magnetized compact
stars in a framework that guarantees the consistency of underlying
compact objects with the currently available astrophysical and
experimental data.
\vspace{12pt}

%\authorcontributions{\hl{Conceptualization, M.S. and A.S.; Methodology, V.B.T.; Supervision, J.J.L. and A.S.; Writing original draft, V.B.T.; Writing review and editing, A.S. All authors have read and agreed to the published version of the~manuscript.}} %mdpi:Please add Author Contributions: For research articles with several authors, a~short paragraph specifying their individual contributions must be~provided.

%\funding{\hl{V.B.T. and M.S. acknowledge the funding
%  support from Science and Engineering Research Board, Department of
%  Science and Technology, Government of India through Project
%  No. EMR/2016/006577 and Ministry of Education, Government of
%  India. M.S. also thanks Alexander von Humboldt Foundation for the support of a visit to Goethe University, Frankfurt am Main.
%  A.S. acknowledges the support through the Deutsche
%  Forschungsgemeinschaft (Grant No. SE 1836/5-1) and European COST
%  Actions ``PHAROS'' (CA16214).}} %mdpi:Please confirm the grant number are right.

\acknowledgments{V.B.T. and M.S. acknowledge the funding
  support from Science and Engineering Research Board, Department of
  Science and Technology, Government of India through Project
  No. EMR/2016/006577 and Ministry of Education, Government of
  India. M.S. also thanks Alexander von Humboldt Foundation for the support of a visit to Goethe University, Frankfurt am Main.
  A.S. acknowledges the support through the Deutsche
  Forschungsgemeinschaft (Grant No. SE 1836/5-1) and European COST
  Actions ``PHAROS'' (CA16214).
  }

%\conflictsofinterest{\hl{The authors declare no conflict of~interest.} }%mdpi:Please confirm

\reftitle{References}


\begin{thebibliography}{999}
\providecommand{\natexlab}[1]{#1}


\bibitem[{Harding} and {Lai}(2006)]{Harding2006}
{Harding}, A.K.; {Lai}, D.
\newblock {Physics of strongly magnetized neutron stars}.
\newblock {\em Rep. Prog. Phys.} {\bf 2006}, {\em
  69},~2631--2708,
  %\href{http://xxx.lanl.gov/abs/astro-ph/0606674}{{\normalfont
  %[arXiv:astro-ph/astro-ph/0606674]}}
    doi:10.1088/0034-4885/69/9/R03.

\bibitem[{Turolla} \em{et~al.}(2015){Turolla}, {Zane}, and
  {Watts}]{Turolla2015RPPh}
{Turolla}, R.; {Zane}, S.; {Watts}, A.L.
\newblock {Magnetars: The physics behind observations. A review}.
\newblock {\em Rep. Prog. Phys.} {\bf 2015}, {\em 78},~116901,
  %\href{http://xxx.lanl.gov/abs/1507.02924}{{\normalfont
  %[arXiv:astro-ph.HE/1507.02924]}}
    doi:10.1088/0034-4885/78/11/116901.

\bibitem[{Margalit} \em{et~al.}(2020){Margalit}, {Beniamini}, {Sridhar}, and
  {Metzger}]{Margalit2020ApJ}
{Margalit}, B.; {Beniamini}, P.; {Sridhar}, N.; {Metzger}, B.D.
\newblock {Implications of a Fast Radio Burst from a Galactic Magnetar}.
\newblock {\em {Astrophys. J. Lett.}} %The journal name we found on Google does not match the one you gave, please confirm.
 {\bf 2020}, {\em 899},~L27,
  %\href{http://xxx.lanl.gov/abs/2005.05283}{{\normalfont
  %[arXiv:astro-ph.HE/2005.05283]}}
    doi:10.3847/2041-8213/abac57.

\bibitem[{Beniamini} \em{et~al.}(2020){Beniamini}, {Wadiasingh}, and
  {Metzger}]{Beniamini2020MNRAS}
{Beniamini}, P.; {Wadiasingh}, Z.; {Metzger}, B.D.
\newblock {Periodicity in recurrent fast radio bursts and the origin of
  ultralong period magnetars}.
\newblock {\em {Mon. Not. R. Astron. Soc.}} {\bf 2020}, {\em 496},~3390--3401,
  %\href{http://xxx.lanl.gov/abs/2003.12509}{{\normalfont
  %[arXiv:astro-ph.HE/2003.12509]}}
    doi:10.1093/mnras/staa1783.

\bibitem[{Beloborodov}(2020)]{Beloborodov2020}
{Beloborodov}, A.M.
\newblock {Blast Waves from Magnetar Flares and Fast Radio Bursts}.
\newblock {\em Astrophys. J.} {\bf 2020}, {\em 896},~142,
  %\href{http://xxx.lanl.gov/abs/1908.07743}{{\normalfont
  %[arXiv:astro-ph.HE/1908.07743]}}
    doi:10.3847/1538-4357/ab83eb.

\bibitem[{Levin} \em{et~al.}(2020){Levin}, {Beloborodov}, and
  {Bransgrove}]{Levin2020ApJ}
{Levin}, Y.; {Beloborodov}, A.M.; {Bransgrove}, A.
\newblock {Precessing Flaring Magnetar as a Source of Repeating FRB
  180916.J0158 + 65}.
\newblock {\em Astrophys. J. Lett.} {\bf 2020}, {\em 895},~L30,
  %\href{http://xxx.lanl.gov/abs/2002.04595}{{\normalfont
  %[arXiv:astro-ph.HE/2002.04595]}}
    doi:10.3847/2041-8213/ab8c4c.

\bibitem[{Zanazzi} and {Lai}(2020)]{Zanazzi2020ApJ}
{Zanazzi}, J.J.; {Lai}, D.
\newblock {Periodic Fast Radio Bursts with Neutron Star Free Precession}.
\newblock {\em Astrophys. J. Lett.} \mbox{{\bf 2020}, {\em 892},~L15,}
  %\href{http://xxx.lanl.gov/abs/2002.05752}{{\normalfont
  %[arXiv:astro-ph.HE/2002.05752]}}
    doi:10.3847/2041-8213/ab7cdd.

\bibitem[{Chakrabarty} \em{et~al.}(1997){Chakrabarty}, {Bandyopadhyay}, and
  {Pal}]{1997PhRvL..78.2898C}
{Chakrabarty}, S.; {Bandyopadhyay}, D.; {Pal}, S.
\newblock {Dense Nuclear Matter in a Strong Magnetic Field}.
\newblock {\em \prl} {\bf 1997}, {\em 78},~2898--2901,
  %\href{http://xxx.lanl.gov/abs/9703034}{{\normalfont
  %[arXiv:astro-ph/9703034]}}
    doi:10.1103/PhysRevLett.78.2898.

\bibitem[{Bandyopadhyay} \em{et~al.}(1997){Bandyopadhyay}, {Chakrabarty}, and
  {Pal}]{1997PhRvL..79.2176B}
{Bandyopadhyay}, D.; {Chakrabarty}, S.; {Pal}, S.
\newblock {Quantizing Magnetic Field and Quark-Hadron Phase Transition in a
  Neutron Star}.
\newblock {\em \prl} {\bf 1997}, {\em 79},~2176--2179,
  %\href{http://xxx.lanl.gov/abs/9703066}{{\normalfont
  %[arXiv:astro-ph/9703066]}}
    doi:10.1103/PhysRevLett.79.2176.

\bibitem[{Broderick} \em{et~al.}(2000){Broderick}, {Prakash}, and
  {Lattimer}]{2000ApJ...537..351B}
{Broderick}, A.; {Prakash}, M.; {Lattimer}, J.M.
\newblock {The Equation of State of Neutron Star Matter in Strong Magnetic
  Fields}.
\newblock {\em Astrophys. J. Lett.} {\bf 2000}, {\em 537},~351--367,
  %\href{http://xxx.lanl.gov/abs/0001537}{{\normalfont
  %[arXiv:astro-ph/0001537]}}
    doi:10.1086/309010.

\bibitem[{Chen} \em{et~al.}(2007){Chen}, {Zhang}, and
  {Liu}]{2007MPLA...22..623C}
{Chen}, W.; {Zhang}, P.Q.; {Liu}, L.G.
\newblock {The Influence of the Magnetic Field on the Properties of Neutron
  Star Matter}.
\newblock {\em Mod. Phys. Lett. A} {\bf 2007}, {\em 22},~623--629,
  %\href{http://xxx.lanl.gov/abs/0505113}{{\normalfont
  %[arXiv:astro-ph/0505113]}}
    doi:10.1142/S0217732307023213.

\bibitem[{Rabhi} \em{et~al.}(2008){Rabhi}, {Provid{\^e}ncia}, and {Da
  Provid{\^e}ncia}]{2008JPhG...35l5201R}
{Rabhi}, A.; {Provid{\^e}ncia}, C.; {Da Provid{\^e}ncia}, J.
\newblock {Stellar matter with a strong magnetic field within density-dependent
  relativistic models}.
\newblock {\em J. Phys. G Nucl. Phys.} {\bf 2008}, {\em
  35},~125201,  %\href{http://xxx.lanl.gov/abs/0810.3390}{{\normalfont
  %[arXiv:nucl-th/0810.3390]}}
    doi:10.1088/0954-3899/35/12/125201.

\bibitem[{Mallick} and {Schramm}(2014)]{2014PhRvC..89d5805M}
{Mallick}, R.; {Schramm}, S.
\newblock {Deformation of a magnetized neutron star}.
\newblock {\em \prc} {\bf 2014}, {\em 89},~045805,
  %\href{http://xxx.lanl.gov/abs/1307.5185}{{\normalfont
  %[arXiv:astro-ph.HE/1307.5185]}}
    doi:10.1103/PhysRevC.89.045805.

\bibitem[{Bonazzola} \em{et~al.}(1993){Bonazzola}, {Gourgoulhon}, {Salgado},
  and {Marck}]{1993A&A...278..421B}
{Bonazzola}, S.; {Gourgoulhon}, E.; {Salgado}, M.; {Marck}, J.A.
\newblock {Axisymmetric rotating relativistic bodies: A new numerical approach
  for `exact' solutions}.
\newblock {\em Astron. Astrophys.} {\bf 1993}, {\em 278},~421--443.

\bibitem[{Bocquet} \em{et~al.}(1995){Bocquet}, {Bonazzola}, {Gourgoulhon}, and
  {Novak}]{1995A&A...301..757B}
{Bocquet}, M.; {Bonazzola}, S.; {Gourgoulhon}, E.; {Novak}, J.
\newblock {Rotating neutron star models with a magnetic field.}
\newblock {\em Astron. Astrophys.} {\bf 1995}, {\em 301},~757,
  %\href{http://xxx.lanl.gov/abs/gr-qc/9503044}{{\normalfont
  %[arXiv:gr-qc/gr-qc/9503044]}}.

\bibitem[{Cardall} \em{et~al.}(2001){Cardall}, {Prakash}, and
  {Lattimer}]{2001ApJ...554..322C}
{Cardall}, C.Y.; {Prakash}, M.; {Lattimer}, J.M.
\newblock {Effects of Strong Magnetic Fields on Neutron Star Structure}.
\newblock {\em Astrophys. J. Lett.} {\bf 2001}, {\em 554},~322--339,
  %\href{http://xxx.lanl.gov/abs/astro-ph/0011148}{{\normalfont
  %[arXiv:astro-ph/astro-ph/0011148]}}
    doi:10.1086/321370.

\bibitem[{Ciolfi} \em{et~al.}(2009){Ciolfi}, {Ferrari}, {Gualtieri}, and
  {Pons}]{2009MNRAS.397..913C}
{Ciolfi}, R.; {Ferrari}, V.; {Gualtieri}, L.; {Pons}, J.A.
\newblock {Relativistic models of magnetars: The twisted torus magnetic field
  configuration}.
\newblock {\em Mon. Not. R. Astron. Soc.} {\bf 2009}, {\em 397},~913--924,
  %\href{http://xxx.lanl.gov/abs/0903.0556}{{\normalfont
  %[arXiv:astro-ph.SR/0903.0556]}}
    doi:10.1111/j.1365-2966.2009.14990.x.

\bibitem[{Ciolfi} \em{et~al.}(2010){Ciolfi}, {Ferrari}, and
  {Gualtieri}]{2010MNRAS.406.2540C}
{Ciolfi}, R.; {Ferrari}, V.; {Gualtieri}, L.
\newblock {Structure and deformations of strongly magnetized neutron stars with
  twisted-torus configurations}.
\newblock {\em Mon. Not. R. Astron. Soc.} {\bf 2010}, {\em 406},~2540--2548,
  %\href{http://xxx.lanl.gov/abs/1003.2148}{{\normalfont
  %[arXiv:astro-ph.SR/1003.2148]}}
    doi:10.1111/j.1365-2966.2010.16847.x.

\bibitem[{Ciolfi} and {Rezzolla}(2013)]{2013MNRAS.435L..43C}
{Ciolfi}, R.; {Rezzolla}, L.
\newblock {Twisted-torus configurations with large toroidal magnetic fields in
  relativistic stars.}
\newblock {\em Mon. Not. R. Astron. Soc.} {\bf 2013}, {\em 435},~L43--L47,
  %\href{http://xxx.lanl.gov/abs/1306.2803}{{\normalfont
  %[arXiv:astro-ph.SR/1306.2803]}}
    doi:10.1093/mnrasl/slt092.

\bibitem[{Chatterjee} \em{et~al.}(2019){Chatterjee}, {Novak}, and
  {Oertel}]{2019PhRvC..99e5811C}
{Chatterjee}, D.; {Novak}, J.; {Oertel}, M.
\newblock {Magnetic field distribution in magnetars}.
\newblock {\em \prc} {\bf 2019}, {\em 99},~055811,
  %\href{http://xxx.lanl.gov/abs/1808.01778}{{\normalfont
  %[arXiv:nucl-th/1808.01778]}}
    doi:10.1103/PhysRevC.99.055811.

\bibitem[{Sinha} and {Sedrakian}(2014)]{Sinha2014}
{Sinha}, M.; {Sedrakian}, A.
\newblock {Upper critical field and (non)-superconductivity of magnetars}.
\newblock {\em arXiv} {\bf 2014}, arXiv:1403.2829.
  %\href{http://xxx.lanl.gov/abs/1403.2829}{{\normalfont
  %[arXiv:astro-ph.SR/1403.2829]}}.

\bibitem[{Sinha} and {Sedrakian}(2015)]{Sinha2015}
{Sinha}, M.; {Sedrakian}, A.
\newblock {Magnetar superconductivity versus magnetism: Neutrino cooling
  processes}.
\newblock {\em \prc} {\bf 2015}, {\em 91},~035805,
  %\href{http://xxx.lanl.gov/abs/1502.02979}{{\normalfont
  %[arXiv:astro-ph.HE/1502.02979]}}
    doi:10.1103/PhysRevC.91.035805.

\bibitem[{Stein} \em{et~al.}(2016){Stein}, {Sedrakian}, {Huang}, and
  {Clark}]{Stein2016}
{Stein}, M.; {Sedrakian}, A.; {Huang}, X.G.; {Clark}, J.W.
\newblock {Spin-polarized neutron matter: Critical unpairing and BCS-BEC
  precursor}.
\newblock {\em \prc} {\bf 2016}, {\em 93},~015802,
  %\href{http://xxx.lanl.gov/abs/1510.06000}{{\normalfont
  %[arXiv:nucl-th/1510.06000]}}
    doi:10.1103/PhysRevC.93.015802.

\bibitem[{Sedrakian} \em{et~al.}(2017){Sedrakian}, {Xu-Guang}, {Sinha}, and
  {Clark}]{Sedrakian2017}
{Sedrakian}, A.; {Xu-Guang}, H.; {Sinha}, M.; {Clark}, J.W.
\newblock {From microphysics to dynamics of magnetars}.
\newblock  {\em J. Phys. Conf. Ser.} \textbf{2017}, \emph{861}, 012025,
  %\href{http://xxx.lanl.gov/abs/1701.00895}{{\normalfont
  %[arXiv:astro-ph.HE/1701.00895]}}
    doi:10.1088/1742-6596/861/1/012025.

\bibitem[{Pons} \em{et~al.}(2009){Pons}, {Miralles}, and
  {Geppert}]{2009A&A...496..207P}
{Pons}, J.A.; {Miralles}, J.A.; {Geppert}, U.
\newblock {Magneto-thermal evolution of neutron stars}.
\newblock {\em Astron. Astrophys.} \mbox{{\bf 2009}, {\em 496},~207--216,}
  %\href{http://xxx.lanl.gov/abs/0812.3018}{{\normalfont
  %[arXiv:astro-ph/0812.3018]}}
    doi:10.1051/0004-6361:200811229.

\bibitem[{Pons} and {Vigan{\`o}}(2019)]{2019LRCA....5....3P}
{Pons}, J.A.; {Vigan{\`o}}, D.
\newblock {Magnetic, thermal and rotational evolution of isolated neutron
  stars}.
\newblock {\em Living Rev. Comput. Astrophys.} {\bf 2019}, {\em
  5},~3,  %\href{http://xxx.lanl.gov/abs/1911.03095}{{\normalfont
  %[arXiv:astro-ph.HE/1911.03095]}}  doi:10.1007/s41115-019-0006-7.

\bibitem[{Sedrakian}(2016)]{Sedrakian2016}
{Sedrakian}, A.
\newblock {Rapid rotational crust-core relaxation in magnetars}.
\newblock {\em Astron. Astrophys.} {\bf 2016}, {\em 587},~L2,
  %\href{http://xxx.lanl.gov/abs/1601.00056}{{\normalfont
  %[arXiv:astro-ph.HE/1601.00056]}}
    doi:10.1051/0004-6361/201628068.

\bibitem[{Hulse} and {Taylor}(1975)]{1975ApJ...195L..51H}
{Hulse}, R.A.; {Taylor}, J.H.
\newblock {Discovery of a pulsar in a binary system.}
\newblock {\em Astrophys. J. Lett.} {\bf 1975}, {\em 195},~L51--L53  doi:10.1086/181708.

\bibitem[{{\''O}zel} \em{et~al.}(2010){{\''O}zel}, {Psaltis}, {Ransom},
  {Demorest}, and {Alford}]{2010ApJ...724L.199O}
{{Ö}zel}, F.; {Psaltis}, D.; {Ransom}, S.; {Demorest}, P.; {Alford}, M.
\newblock {The Massive Pulsar PSR J1614-2230: Linking Quantum Chromodynamics,
  Gamma-ray Bursts, and Gravitational Wave Astronomy}.
\newblock {\em Astrophys. J. Lett.} {\bf 2010}, {\em 724},~L199--L202,
  %\href{http://xxx.lanl.gov/abs/1010.5790}{{\normalfont
  %[arXiv:astro-ph.HE/1010.5790]}}
    doi:10.1088/2041-8205/724/2/L199.

\bibitem[{Antoniadis} \em{et~al.}(2013){Antoniadis}, {Freire}, {Wex}, {Tauris},
  {Lynch}, {van Kerkwijk}, {Kramer}, {Bassa}, {Dhillon}, {Driebe}, {Hessels},
  {Kaspi}, {Kondratiev}, {Langer}, {Marsh}, {McLaughlin}, {Pennucci}, {Ransom},
  {Stairs}, {van Leeuwen}, {Verbiest}, and {Whelan}]{2013Sci...340..448A}
  
{Antoniadis}, J.; {Freire}, P.C.C.; {Wex}, N.; {Tauris}, T.M.; {Lynch}, R.S.;
  {van Kerkwijk}, M.H.; {Kramer}, M.; {Bassa}, C.; {Dhillon}, V.S.; {Driebe},
  T.; et al. 
\newblock {A Massive Pulsar in a Compact Relativistic Binary}.
\newblock {\em Science} {\bf 2013}, {\em 340},~448,
  %\href{http://xxx.lanl.gov/abs/1304.6875}{{\normalfont
  %[arXiv:astro-ph.HE/1304.6875]}}
    doi:10.1126/science.1233232.

\bibitem[{Cromartie} \em{et~al.}(2020){Cromartie}, {Fonseca}, {Ransom},
  {Demorest}, {Arzoumanian}, {Blumer}, {Brook}, {DeCesar}, {Dolch}, {Ellis},
  {Ferdman}, {Ferrara}, {Garver-Daniels}, {Gentile}, {Jones}, {Lam}, {Lorimer},
  {Lynch}, {McLaughlin}, {Ng}, {Nice}, {Pennucci}, {Spiewak}, {Stairs},
  {Stovall}, {Swiggum}, and {Zhu}]{2020NatAs...4...72C}
  
{Cromartie}, H.T.; {Fonseca}, E.; {Ransom}, S.M.; {Demorest}, P.B.;
  {Arzoumanian}, Z.; {Blumer}, H.; {Brook}, P.R.; {DeCesar}, M.E.; {Dolch}, T.;
  {Ellis}, J.A.; et al. 
\newblock {Relativistic Shapiro delay measurements of an extremely massive
  millisecond pulsar}.
\newblock {\em Nat. Astron.} {\bf 2020}, {\em 4},~72--76,
  %\href{http://xxx.lanl.gov/abs/1904.06759}{{\normalfont
  %[arXiv:astro-ph.HE/1904.06759]}}
    doi:10.1038/s41550-019-0880-2.

\bibitem[{Abbott} \em{et~al.}(2020){Abbott}, {Abbott}, {Abraham}, {Acernese},
  {Ackley}, {Adams}, {Adhikari}, {Adya}, {Affeldt}, {Agathos}, and
  et~al.]{2020ApJ...896L..44A}
  
{Abbott}, R.; {Abbott}, T.D.; {Abraham}, S.; {Acernese}, F.; {Ackley}, K.;
  {Adams}, C.; {Adhikari}, R.X.; {Adya}, V.B.; {Affeldt},~C.; {Agathos}, M.;
  et~al.
\newblock {GW190814: Gravitational Waves from the Coalescence of a 23 Solar
  Mass Black Hole with a 2.6 Solar Mass Compact Object}.
\newblock {\em Astrophys. J. Lett.} {\bf 2020}, {\em 896},~L44,
  %\href{http://xxx.lanl.gov/abs/2006.12611}{{\normalfont
  %[arXiv:astro-ph.HE/2006.12611]}}
    doi:10.3847/2041-8213/ab960f.

\bibitem[{Miller} \em{et~al.}(2019){Miller}, {Lamb}, {Dittmann}, {Bogdanov},
  {Arzoumanian}, {Gendreau}, {Guillot}, {Harding}, {Ho}, {Lattimer}, {Ludlam},
  {Mahmoodifar}, {Morsink}, {Ray}, {Strohmayer}, {Wood}, {Enoto}, {Foster},
  {Okajima}, {Prigozhin}, and {Soong}]{2019ApJ...887L..24M}
  
{Miller}, M.C.; {Lamb}, F.K.; {Dittmann}, A.J.; {Bogdanov}, S.; {Arzoumanian},
  Z.; {Gendreau}, K.C.; {Guillot}, S.; {Harding},~A.K.; {Ho}, W.C.G.;
  {Lattimer}, J.M.; et al.
\newblock {PSR J0030+0451 Mass and Radius from NICER Data and Implications for
  the Properties of Neutron Star Matter}.
\newblock {\em Astrophys. J. Lett.} {\bf 2019}, {\em 887},~L24,
  %\href{http://xxx.lanl.gov/abs/1912.05705}{{\normalfont
  %[arXiv:astro-ph.HE/1912.05705]}}
    doi:10.3847/2041-8213/ab50c5.

\bibitem[{Riley} \em{et~al.}(2019){Riley}, {Watts}, {Bogdanov}, {Ray},
  {Ludlam}, {Guillot}, {Arzoumanian}, {Baker}, {Bilous}, {Chakrabarty},
  {Gendreau}, {Harding}, {Ho}, {Lattimer}, {Morsink}, and
  {Strohmayer}]{2019ApJ...887L..21R}
  
{Riley}, T.E.; {Watts}, A.L.; {Bogdanov}, S.; {Ray}, P.S.; {Ludlam}, R.M.;
  {Guillot}, S.; {Arzoumanian}, Z.; {Baker}, C.L.; {Bilous}, A.V.;
  {Chakrabarty}, D.; et al. 
\newblock {A NICER View of PSR J0030+0451: Millisecond Pulsar Parameter
  Estimation}.
\newblock {\em Astrophys. J. Lett.} {\bf 2019}, {\em 887},~L21,
  %\href{http://xxx.lanl.gov/abs/1912.05702}{{\normalfont
  %[arXiv:astro-ph.HE/1912.05702]}}
    doi:10.3847/2041-8213/ab481c.

\bibitem[{Lenske} and {Fuchs}(1995)]{1995PhLB..345..355L}
{Lenske}, H.; {Fuchs}, C.
\newblock {Rearrangement in the density dependent relativistic field theory of
  nuclei}.
\newblock {\em \mbox{Phys. Lett. B}} {\bf 1995}, {\em 345},~355--360  doi:10.1016/0370-2693(94)01664-X.

\bibitem[{Fuchs} \em{et~al.}(1995){Fuchs}, {Lenske}, and
  {Wolter}]{1995PhRvC..52.3043F}
{Fuchs}, C.; {Lenske}, H.; {Wolter}, H.H.
\newblock {Density dependent hadron field theory}.
\newblock {\em \prc} {\bf 1995}, {\em 52},~3043--3060,
  %\href{http://xxx.lanl.gov/abs/nucl-th/9507044}{{\normalfont
  %[arXiv:nucl-th/nucl-th/9507044]}}
    doi:10.1103/PhysRevC.52.3043.

\bibitem[{Typel} and {Wolter}(1999)]{1999NuPhA.656..331T}
{Typel}, S.; {Wolter}, H.H.
\newblock {Relativistic mean field calculations with density-dependent
  meson-nucleon coupling}.
\newblock {\em \nphysa} {\bf 1999}, {\em 656},~331--364  doi:10.1016/S0375-9474(99)00310-3.

\bibitem[{Hofmann} \em{et~al.}(2001){Hofmann}, {Keil}, and
  {Lenske}]{2001PhRvC..64b5804H}
{Hofmann}, F.; {Keil}, C.M.; {Lenske}, H.
\newblock {Application of the density dependent hadron field theory to neutron
  star matter}.
\newblock {\em \prc} {\bf 2001}, {\em 64},~025804,
  %\href{http://xxx.lanl.gov/abs/nucl-th/0008038}{{\normalfont
  %[arXiv:nucl-th/nucl-th/0008038]}}
    doi:10.1103/PhysRevC.64.025804.

\bibitem[{Li} \em{et~al.}(2008){Li}, {Chen}, and {Ko}]{2008PhR...464..113L}
{Li}, B.A.; {Chen}, L.W.; {Ko}, C.M.
\newblock {Recent progress and new challenges in isospin physics with heavy-ion
  reactions}.
\newblock {\em \physrep} {\bf 2008}, {\em 464},~113--281,
  %\href{http://xxx.lanl.gov/abs/0804.3580}{{\normalfont
  %[arXiv:nucl-th/0804.3580]}}
    doi:10.1016/j.physrep.2008.04.005.

\bibitem[{Drago} \em{et~al.}(2014{\natexlab{a}}){Drago}, {Lavagno}, and
  {Pagliara}]{2014PhRvD..89d3014D}
{Drago}, A.; {Lavagno}, A.; {Pagliara}, G.
\newblock {Can very compact and very massive neutron stars both exist?}
\newblock {\em \prd} {\bf 2014}, {\em 89},~043014,
  %\href{http://xxx.lanl.gov/abs/1309.7263}{{\normalfont
  %[arXiv:nucl-th/1309.7263]}}
    doi:10.1103/PhysRevD.89.043014.

\bibitem[{Drago} \em{et~al.}(2014{\natexlab{b}}){Drago}, {Lavagno}, {Pagliara},
  and {Pigato}]{2014PhRvC..90f5809D}
{Drago}, A.; {Lavagno}, A.; {Pagliara}, G.; {Pigato}, D.
\newblock {Early appearance of {\ensuremath{\Delta}} isobars in neutron stars}.
\newblock {\em \prc} {\bf 2014}, {\em 90},~065809  doi:10.1103/PhysRevC.90.065809.

\bibitem[{Cai} \em{et~al.}(2015){Cai}, {Fattoyev}, {Li}, and
  {Newton}]{2015PhRvC..92a5802C}
{Cai}, B.J.; {Fattoyev}, F.J.; {Li}, B.A.; {Newton}, W.G.
\newblock {Critical density and impact of {\ensuremath{\Delta}} (1232 )
  resonance formation in neutron stars}.
\newblock {\em \prc} {\bf 2015}, {\em 92},~015802,
  %\href{http://xxx.lanl.gov/abs/1501.01680}{{\normalfont [1501.01680]}}
    doi:10.1103/PhysRevC.92.015802.

\bibitem[{Glendenning}(1985)]{1985ApJ...293..470G}
{Glendenning}, N.K.
\newblock {Neutron stars are giant hypernuclei?}
\newblock {\em Astrophys. J.} %We can’t find relevant information on Google, please complete the abbreviation of the journal name.
 {\bf 1985}, {\em 293},~470--493,  doi:10.1086/163253.

\bibitem[{Li} \em{et~al.}(2018){Li}, {Sedrakian}, and
  {Weber}]{2018PhLB..783..234L}
{Li}, J.J.; {Sedrakian}, A.; {Weber}, F.
\newblock {Competition between delta isobars and hyperons and properties of
  compact stars}.
\newblock {\em Phys. Lett. B} {\bf 2018}, {\em 783},~234--240,
  %\href{http://xxx.lanl.gov/abs/1803.03661}{{\normalfont
  %[arXiv:nucl-th/1803.03661]}}
    doi:10.1016/j.physletb.2018.06.051.

\bibitem[Li and Sedrakian(2019)]{Li2019ApJ}
Li, J.J.; Sedrakian, A.
\newblock {Implications from GW170817 for $\Delta$-isobar Admixed Hypernuclear
  Compact Stars}.
\newblock {\em Astrophys. J. Lett.} {\bf 2019}, {\em 874},~L22,
  %\href{http://xxx.lanl.gov/abs/1904.02006}{{\normalfont
  %[arXiv:nucl-th/1904.02006]}}
    doi:10.3847/2041-8213/ab1090.

\bibitem[{Sedrakian} \em{et~al.}(2020){Sedrakian}, {Weber}, and
  {Li}]{2020arXiv200709683S}
{Sedrakian}, A.; {Weber}, F.; {Li}, J.J.
\newblock {Confronting GW190814 with hyperonization in dense matter and
  hypernuclear compact stars}.
\newblock {\em arXiv} {\bf 2020}, arXiv:2007.09683.
  %\href{http://xxx.lanl.gov/abs/2007.09683}{{\normalfont
  %[arXiv:astro-ph.HE/2007.09683]}}.

\bibitem[{Li} and {Sedrakian}(2019)]{2019PhRvC.100a5809L}
{Li}, J.J.; {Sedrakian}, A.
\newblock {Constraining compact star properties with nuclear saturation
  parameters}.
\newblock {\em \prc} {\bf 2019}, {\em 100},~015809,
  %\href{http://xxx.lanl.gov/abs/1903.06057}{{\normalfont
  %[arXiv:astro-ph.HE/1903.06057]}}
    doi:10.1103/PhysRevC.100.015809.

\bibitem[{Tolos} \em{et~al.}(2017){Tolos}, {Centelles}, and
  {Ramos}]{2017PASA...34...65T}
{Tolos}, L.; {Centelles}, M.; {Ramos}, A.
\newblock {The Equation of State for the Nucleonic and Hyperonic Core of
  Neutron Stars}.
\newblock {\em  Publ. Astron. Soc. Aust.} {\bf 2017}, {\em 34},~e065,
  %\href{http://xxx.lanl.gov/abs/1708.08681}{{\normalfont
  %[arXiv:astro-ph.HE/1708.08681]}}
    doi:10.1017/pasa.2017.60.

\bibitem[{Sinha} \em{et~al.}(2013){Sinha}, {Mukhopadhyay}, and
  {Sedrakian}]{2013NuPhA.898...43S}
{Sinha}, M.; {Mukhopadhyay}, B.; {Sedrakian}, A.
\newblock {Hypernuclear matter in strong magnetic field}.
\newblock {\em \nphysa} {\bf 2013}, {\em 898},~43--58,
  %\href{http://xxx.lanl.gov/abs/1005.4995}{{\normalfont
  %[arXiv:astro-ph.HE/1005.4995]}}
    doi:10.1016/j.nuclphysa.2012.12.076.

\bibitem[{de Paoli} \em{et~al.}(2013){de Paoli}, {Castro}, {Menezes}, and
  {Barros}]{2013JPhG...40e5007D}
{De Paoli}, M.G.; {Castro}, L.B.; {Menezes}, D.P.; {Barros}, C.C.J.
\newblock {Rarita-Schwinger particles under the influence of strong magnetic
  fields}.
\newblock {\em J. Phys. G Nucl. Phys.} {\bf 2013}, {\em
  40},~055007,  %\href{http://xxx.lanl.gov/abs/1207.4063}{{\normalfont
  %[arXiv:math-ph/1207.4063]}}
    doi:10.1088/0954-3899/40/5/055007.

\bibitem[{Lalazissis} \em{et~al.}(2005){Lalazissis}, {Nik{\v{s}}i{\'c}},
  {Vretenar}, and {Ring}]{2005PhRvC..71b4312L}
{Lalazissis}, G.A.; {Nik{\v{s}}i{\'c}}, T.; {Vretenar}, D.; {Ring}, P.
\newblock {New relativistic mean-field interaction with density-dependent
  meson-nucleon couplings}.
\newblock {\em \prc} {\bf 2005}, {\em 71},~024312  doi:10.1103/PhysRevC.71.024312.

\bibitem[{Banik} and {Bandyopadhyay}(2001)]{2001PhRvC..64e5805B}
{Banik}, S.; {Bandyopadhyay}, D.
\newblock {Third family of superdense stars in the presence of antikaon
  condensates}.
\newblock {\em \prc} {\bf 2001}, {\em 64},~055805,
  %\href{http://xxx.lanl.gov/abs/0106406}{{\normalfont
  %[arXiv:astro-ph/0106406]}}
    doi:10.1103/PhysRevC.64.055805.

\bibitem[{Li} \em{et~al.}(2018){Li}, {Long}, and
  {Sedrakian}]{2018EPJA...54..133L}
{Li}, J.J.; {Long}, W.H.; {Sedrakian}, A.
\newblock {Hypernuclear stars from relativistic Hartree-Fock density functional
  theory}.
\newblock {\em Eur. Phys. J. A} {\bf 2018}, {\em 54},~133,
  %\href{http://xxx.lanl.gov/abs/1801.07084}{{\normalfont
  %[arXiv:nucl-th/1801.07084]}}
    doi:10.1140/epja/i2018-12566-6.

\end{thebibliography}
\end{document}